\documentclass[3p,times]{elsarticle}

\usepackage{ecrc}
\usepackage{epsfig}
\usepackage{caption}
\usepackage{subfig}
\usepackage{titlesec}
\usepackage{braket}
\usepackage{amsmath}
\usepackage{longtable}
\usepackage{pdflscape}
\usepackage{hyperref}
\usepackage{color}
\usepackage{hyperref} 

\volume{00}

\firstpage{1}

\journalname{Computer Science Review}

\runauth{}

\jid{Comput Sci Rev}

\jnltitlelogo{Computer Science Review}

\CopyrightLine{2022}{Published by Elsevier Ltd.}

\usepackage{amssymb}

\usepackage[figuresright]{rotating}

\begin{document}

\begin{frontmatter}



\dochead{Review article}
\title{Systematic Literature Review: Quantum Machine Learning and its applications}


\author[esalab]{David Peral García}
\author[ibmquantum]{Juan Cruz-Benito}
\author[grial]{Francisco José García-Peñalvo}

\address[esalab]{Expert Systems and Applications Laboratory - ESALAB, Faculty of Science, University of Salamanca, Plaza de los Caídos s/n, 37008 Salamanca, Spain}
\address[ibmquantum]{IBM Quantum, IBM T.J. Watson Research Center, Yorktown Heights, NY 10598, USA}
\address[grial]{GRIAL Research Group, Department of Computers and Automatics, Research Institute for Educational Sciences, University of Salamanca, Paseo de Canalejas, 169, Salamanca 37008, Spain}

\begin{abstract}
Quantum physics has changed the way we understand our environment, and one of its branches, quantum mechanics, has demonstrated accurate and consistent theoretical results. Quantum computing is the process of performing calculations using quantum mechanics. This field studies the quantum behavior of certain subatomic particles (photons, electrons, etc.) for subsequent use in performing calculations, as well as for large-scale information processing. These advantages are achieved through the use of quantum features, such as entanglement or superposition.
These capabilities can give quantum computers an advantage in terms of computational time and cost over classical computers. Nowadays, scientific challenges are impossible to perform by classical computation due to computational complexity (more bytes than atoms in the observable universe) or the time it would take (thousands of years), and quantum computation is the only known answer. However, current quantum devices do not have yet the necessary qubits and are not fault-tolerant enough to achieve these goals. Nonetheless, there are other fields like machine learning, finance, or chemistry where quantum computation could be useful with current quantum devices. This manuscript aims to present a review of the literature published between 2017 and 2023 to identify, analyze, and classify the different types of algorithms used in quantum machine learning and their applications. The methodology follows the guidelines related to Systematic Literature Review methods, such as the one proposed by Kitchenham and other authors in the software engineering field.
Consequently, this study identified 94 articles that used quantum machine learning techniques and algorithms and shows their implementation using computational quantum circuits or \textit{ansatzs}. The main types of found algorithms are quantum implementations of classical machine learning algorithms, such as support vector machines or the k-nearest neighbor model, and classical deep learning algorithms, like quantum neural networks. One of the most relevant applications in the machine learning field is image classification.
Many articles, especially within the classification, try to solve problems currently answered by classical machine learning but using quantum devices and algorithms. Even though results are promising, quantum machine learning is far from achieving its full potential. An improvement in quantum hardware is required for this potential to be achieved since the existing quantum computers lack enough quality, speed, and scale to allow quantum computing to achieve its full potential. \cite{li2021quantum}
\end{abstract}


\end{frontmatter}


\section{Introduction}
\label{}

Currently, there is an ongoing challenge in the scientific world to create quantum computers capable of substantial advantages for certain problems compared to conventional computers. The automated tools improvement and methods to assist in the simulation and design of the corresponding applications are required along with the device's development. Otherwise, one could end up in a situation where one has powerful quantum computers but very few adequate means to use them \cite{Zulehner2020}. On the other hand, scientific progress in materials creation, hardware fabrication, and disciplines such as error correction and compilation have enabled us to create large-scale and increasingly fault-tolerant quantum computers \cite{Benedetti2019}. Some of the main objectives to be achieved are improving simulation times in chemical compounds, leading to substantial reductions in drug creation times. Also, the creation of complex cryptographic systems will allow the creation of secure computers, guaranteeing internet security for all users and the design of new artificial intelligence algorithms currently housed in most devices, from better prediction and recommendation systems to new industrial ones support systems.

Since the 90s, we have observed the proposal and development of various highly relevant algorithms that demonstrate the computational advantage over classical computation. Some of them are Grover's search algorithm \cite{Grover1996} and Shor's algorithm \cite{Shor1997} for integer factorization in polynomial time, both of which significantly outperform conventional machines. Recently, the application area of quantum algorithms has expanded dramatically and provided efficient methods in areas, such as chemistry \cite{Cao2019, McArdle2020, Bauer2020}, communications \cite{ASHWIN2023108565}, solving systems of linear equations \cite{Harrow2009}, physics simulations \cite{Nachman2021, bauer2021simulating}, cybersecurity \cite{Bennett_2014, Scarani_2009, Zhang_2017} and machine learning \cite{Biamonte2017, Schuld2014}. One of the most emerging areas has been the machine learning field. With existing quantum devices and algorithms, quantum algorithms have already improved some classical processes, and, in contrast, classical machine learning is used to enhance quantum procedures.

Within the application of classical machine learning techniques for improvement of the quantum world, recent studies show the detection of quantum entanglement with unsupervised training in fully and partially entangled structures \cite{chen2021detecting}. Furthermore, deep learning techniques allow us to reduce the noise produced in quantum systems \cite{ai2021experimentally} or determine the structural properties and molecular dynamics in the quantum scenario \cite{domingo2021deep}. Regarding the design and implementation of hybrid deep learning algorithms, the use of hybrid graphical convolutional networks (QGCNN) has been proposed, which, compared to quantum convolutional neural networks, the classical multilayer perceptron (MLP) and classical convolutional networks, can achieve better performances, for example, in high-energy physics (HEP) \cite{chen2021detecting}. In the finance application field, hybrid neural networks have been used for predicting Gross Domestic Product growth \cite{Alaminos2021}. In the healthcare field, the use of Boltzmann machines for classifying patients with lung cancer \cite{Jain2020}.

Bringing all these ideas together, the topic of this article is the link between the different types of quantum machine learning algorithms and their applications: What types of algorithms are used? Where are these algorithms implemented? We will address these questions based on the literature published in these research fields. To achieve these objectives, we conducted a systematic review of the literature. This paper is organized as follows: section 2 presents the systematic review protocol, including its different aspects and necessary steps. Section 3 presents the systematic review results, section 4 discusses the topic, and section 5 presents the conclusions.

\section{Materials and methods}

To identify, evaluate, and interpret the studies available in the literature and taking into account the research questions proposed by the authors, the methodology proposed by Kitchenham et al. \cite{kitchenham2007guidelines} is used, thus organizing it into planning, conducting, and reporting the study. The structure of this section will be distributed according to \cite{CRUZBENITO2018, holgado2020, Penalvo_2022}. First, the databases will be selected, then the research questions will be defined as well as the inclusion and exclusion criteria and the definition of the search string, and finally, the review process will be carried out.

While working on this systematic literature review, we used Parsifal (\url{https://parsif.al/}) to manage traceability and Mendeley (\url{https://www.mendeley.com/}) as the bibliographic manager. On the other hand, two bibliographic reviews have been taken as a reference to structure this study, one of them from a general perspective of quantum computing \cite{OQuinn2020} and another one from a more specific context of quantum machine learning \cite{Benedetti2019}. Furthermore, to facilitate the reproducibility of this study, we provide a public record of the review procedure and steps in \url{https://github.com/DaveHub5/SLR_Quantum_Machine_Learning}. There, the readers can observe the reviewed articles' status through the PRISMA diagram's different phases \cite{MOHER2010336}.

\begin{figure}[!htbp]
\centering
\includegraphics[width=0.85\textwidth]{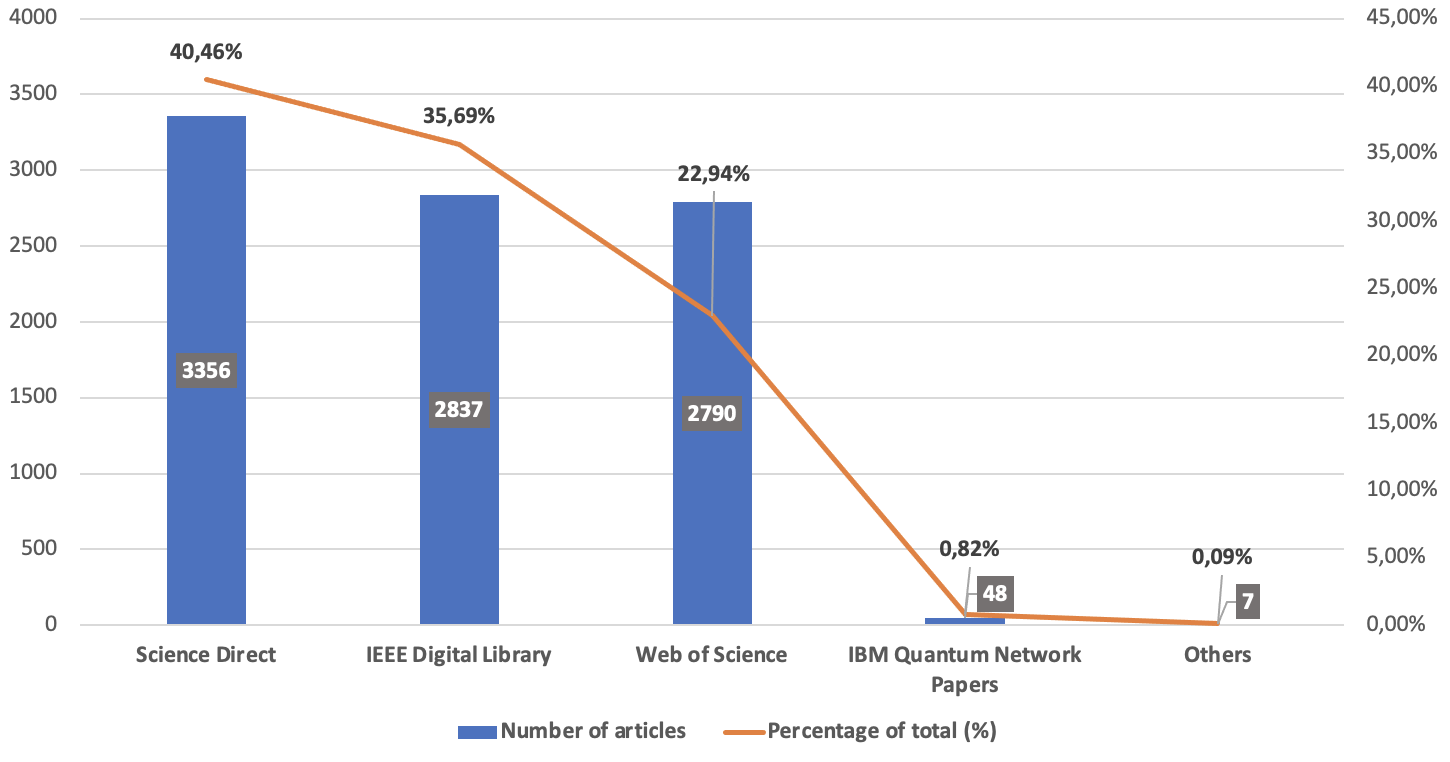}
\caption{Percentage of Articles by Source}
\label{fig:poocentaje_Articulos}
\end{figure}

\begin{figure}[!htbp]
\centering
\includegraphics[width=0.85\textwidth]{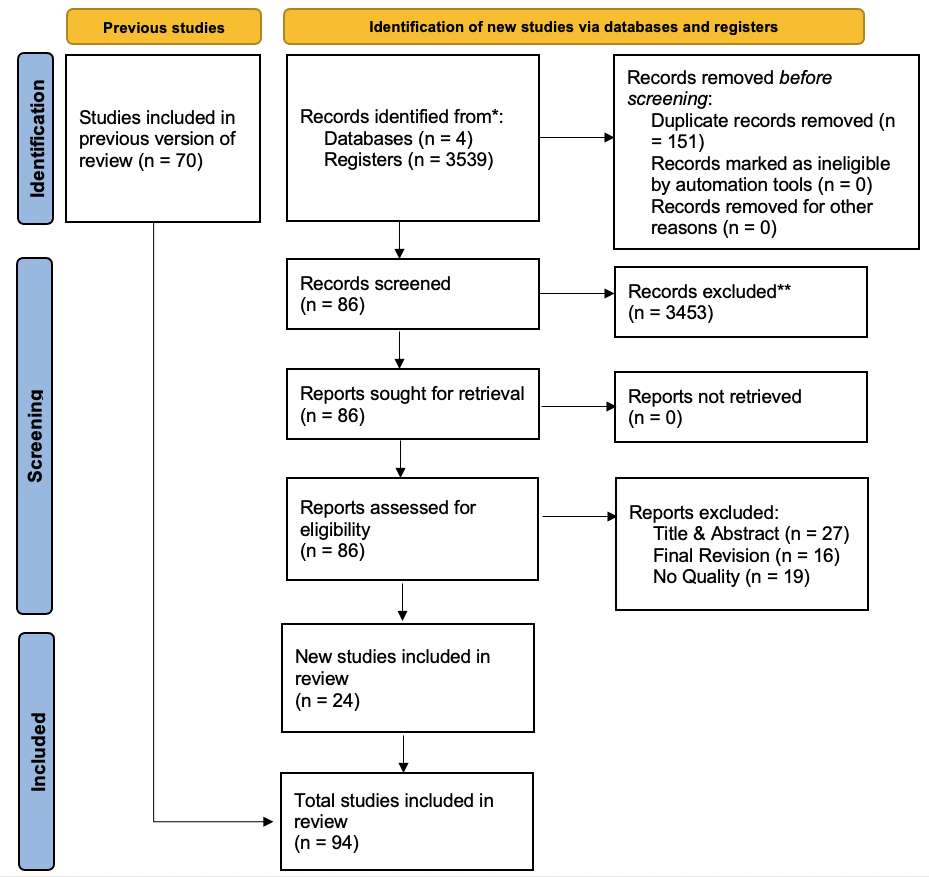}
\caption{PRISMA Diagram}
\label{fig:PRISMA}
\end{figure}

We collected in total 5497 articles [fig:\ref{fig:poocentaje_Articulos}], 1962 obtained from IEEE Digital Library, 1261 from Web of Science, 2224 from Science Direct, 45 from IBM Quantum Network Papers, and 5 articles from other sources published between 2017 and 2022. IBM Quantum Network Papers is an index maintained by IBM \url{https://airtable.com/shr5QnbLgraHRPx35/tblqDKDgMVdH6YGSE}. The query strings used for the search process were "quantum machine learning", "quantum deep learning", or "quantum neural network*". At the end of the study, we identified 94 articles that met the inclusion criteria, as we explain below. A summary of the identification and selection of the articles included in this systematic literature review can be seen in the PRISMA diagram [fig:\ref{fig:PRISMA}]. 

\subsection{Review planning}
This review attempts to show solutions and improvements for the development of quantum circuit simulation tools, quantum circuit design, and the creation and adaptation of different algorithms for quantum computation. The paper aims to serve as an introduction to this emerging field, along with a discussion of recent advances and presenting the problems that remain to be solved \cite{OQuinn2020}.

The field of quantum computing is a recent field compared with classical computing. For this reason, we will focus on the advances and applications of the last six years (2017-2023). We will compare the different techniques used as well as the challenges. The journals used for the articles retrieval have been IEEE Digital Library, Web of Science, and Science Direct. Also, an index of articles related to the cited topic has been included. These databases were selected according to the following requirements: the database is general-purpose journals with a large number of articles, the database admits full-length searches or searches only in specific fields of the works, is one of the most relevant in the research area of computer science, and the database is available for us, the licenses held by the University of Salamanca. IBM Quantum Network Papers is open access to everyone but is notably different, it is an article collection that includes particular articles of interest to the company or includes work that uses IBM Quantum hardware and tools. We include this not-indexed database with ensured contrasted quality content because essential articles are not included in the other databases.

\subsection{Research questions}
This article will focus on and try to answer the following questions.

\begin{itemize}
    \setlength\itemsep{0em}
    \item RQ1: What type of algorithm is used?	
    \item RQ2: Do they use hybrid or pure quantum algorithms?
    \item RQ3: What are the application domains?
    \item RQ4: What are the metrics used to measure the algorithm's success rate or solution proposed? What are the metric values reported?
    \item RQ5: Are the experiments carried out on simulators or real quantum devices?
\end{itemize}

After the research questions are defined, we used the PICOC method proposed by \cite{Petticrew2006} to define our review scope:

\begin{itemize}
    \setlength\itemsep{0em}
    \item Population (P): The target group for the investigation. In this study: Quantum Machine Learning.
    \item Intervention (I): specifies the investigation aspects or issues of interest for the researchers. In our case, those aspects or issues provide support or analyze QML processes.
    \item Comparison (C): the aspect of the investigation to which the intervention is being compared. No comparison intervention has been planned in our study.
    \item Outcomes (O): the effect of the intervention. We are looking for Quantum Machine Learning improvements and implementations.
    \item Context(C): the setting or environment of the investigation. In our case, they are those environments related to QML (in the industry, universities, etc.).
\end{itemize}

\subsection{Inclusion and exclusion criteria}
Once these questions have been defined, the following inclusion and exclusion criteria are defined.

Inclusion criteria:

\begin{itemize}
    \setlength\itemsep{0em}
    \item IC1: The solutions provided by the articles are either applied in software or can be simulated or implemented in simulators or quantum devices AND
    \item IC2: The papers are in a Journal, Book, Preprint or Conference AND
    \item IC3: The papers are written in English or Spanish AND
    \item IC4: The papers provide a practical solution
\end{itemize}

Exclusion criteria:
\begin{itemize}
    \setlength\itemsep{0em}
    \item EC1: The solutions provided by the articles are not applied in software and can not be simulated or implemented in simulators or quantum devices OR
    \item EC2: The papers are not in a Journal, Book, Preprint or Conference  OR
    \item EC3: The papers are not written in English or Spanish OR
    \item EC4: The papers do not provide a practical solution
\end{itemize}

\subsection{Query string}
The first step is to collect all the articles from 2017 to 2023, we have selected this period because it corresponds to the beginning of the rise and growth of quantum devices, as well as the proliferation and possible use of noisy quantum computers or noisy intermediate-scale quantum (NISQ) devices \cite{Preskill2018quantumcomputingin}. To carry out this process, we define some keywords as follows: \textit{quantum deep learning}, \textit{quantum machine learning}, \textit{quantum neural network}.

For this purpose, we use the following queries:

\begin{itemize}
    \setlength\itemsep{0em}
    \item IEEE Digital Library: (All Metadata : quantum \textit{machine learning}) OR (All Metadata:quantum deep learning) OR (All Metadata : quantum neural network*)
    \\Filters Applied: 2017 - 2023
    \item Web of Science: (quantum \textit{machine learning} OR quantum deep learning OR quantum neural network*)
    \\Time period: 2017-2022
    \item Science Direct: quantum \textit{machine learning} or quantum deep learning or quantum neural network
    \\From 2017 to 2022
\end{itemize}

We observe no possible query in the case of IBM Quantum Network Papers. However, we applied a filter in the field \textit{Research Domain} of the type \textit{Machine Learning} because the index of articles is specific to the topic of quantum computing, so there is no need to indicate the word quantum itself.

\subsection{Review process}
When exporting the information from the articles, we encountered a problem in the case of IBM Quantum Network Papers, which does not allow export to BibTeX format, although it has a link to the arvix.org page.

From these queries, we obtained 5497 articles, 1962 from IEEE Digital Library, 1261 from Web of Science, 2224 from Science Direct, 45 from IBM Quantum Network Papers, and 5 articles from other sources. After removing duplicates, we got 1583 unique records. We searched through the titles for those referring to the following words \textit{quantum machine learning}, \textit{quantum deep learning}, \textit{quantum neural network}, leaving us with 382. With this manageable number, we proceeded to read the abstracts, obtaining 338 articles. With a more exhaustive review, we reduce the list to 163, to which we will apply quality criteria. These criteria are as follows:

\begin{itemize}
    \setlength\itemsep{0em}
    \item QC1. Are the research aims specified?
    \item QC2. Is the solution implemented?
    \item QC3. Does the study provide accurate results?
    \item QC4. Does the study show the applied quantum circuit used?
\end{itemize}
We apply these quality criteria for every paper. Each article should have a score $>$= 3.0 out of 4.0 points to be selected, considering one point for each item in the quality criteria list. We obtained a set of 70 articles as the final result.

\section{Results}
\subsection{Algorithm types}
Regarding RQ1, \hyperref[tab:Tabla3Columnas]{Table 1} show the distribution of algorithms and the associated papers. In order to answer RQ2, in this section, we will focus on the different types of Quantum Machine Learning algorithms. We will give a brief explanation of their mathematical basis and implementation of each one of them. The first step of the following algorithms is data encoding, i.e., transforming bits into qubits, a type of information understandable by quantum computers. The second is the creation of the quantum circuit or ansatz, in which the different quantum logic gates are applied to the states. Finally, the reading and measurement of the qubits.

\begin{sidewaystable}
\begin{longtable}
		{|p{2,5cm}  p{0,3cm}  p{0,3cm}  p{0,3cm}  p{0,3cm}  p{0,3cm}  p{0,3cm}  p{0,3cm}  p{0,3cm} p{0,3cm}  p{0,3cm}  p{0,3cm} p{0,3cm}  p{0,3cm}  p{0,3cm} p{0,3cm}  p{0,3cm}  p{0,3cm} p{0,3cm}  p{0,3cm}  p{0,3cm} p{0,3cm}  p{0,3cm} p{0,3cm} p{0,3cm} p{0,3cm}|} \hline
		   \textbf{Algorithm} & 

		   \cite{Huang2020} & 
		   \cite{Srikumar_2022} & 
		   \cite{suzuki2021natural} &
		   \cite{Chalumuri2021} & 
		   \cite{wu2021application} & 
		   \cite{Chen2020b} & \cite{Tacchino_2021} & \cite{Bausch2018} &
		   \cite{LaBorde2020} & 
		   \cite{jia2019orthogonal} & \cite{wang2020orthogonal} & \cite{kerenidis2021classical} &
		   \cite{Konar2021} & 
		   \cite{Lukac2018} & 
		   \cite{Li2020b} & 
		   \cite{Chen2020b} & 
		   \cite{Tacchino2020} &
		   \cite{Wang2021} & \cite{Goncalves2017} &
		   \cite{Situ2020a} & \cite{Liu2021} & 
              \cite{Anand2021} &
              \cite{Ceschini_2022} & 
              \cite{Hong2023c} &
              \cite{CHEN2023321} \\ \hline

            \textbf{vVQC}                  & X &   &   & X &   &   &   &   &   &   &   &   &   &   &   &   &   &   &   &   &   &   &  &  &  X  \\ \hline
            \textbf{HQA}                    &   & X &   &   &   &   &   &   &   &   &   &   &   &   &   &   &   &   &   &   &   &   &  &  &   \\ \hline
            \textbf{QRC}                    &   &   & X &   &   &   &   &   &   &   &   &   &   &   &   &   &   &   &   &   &  &   &   &  &   \\ \hline
            \textbf{QMCC}                   &   &   &   & X &   &   &   &   &   &   &   &   &   &   &   &   &   &   &   &   &   &   &   &  &   \\ \hline
            \textbf{QSVM-Kernel}            &   &   &   &   & X &   &   &   &   &   &   &   &   &   &   &   &   &   &   &   &   &   &  &  &   \\ \hline
            \textbf{RQNN}                   &   &   &   &   &   & X & X & X &   &   &   &   &   &   &   &   &   &   &   &   &   &   &  & &    \\ \hline
            \textbf{HKNN}                  &   &   &   &   &   &   &   &   & X &   &   &   &   &   &   &   &   &   &   &   &   &   &   & &    \\ \hline
            \textbf{OrthogonalNN}           &   &   &   &   &   &   &   &   &   & X & X & X &   &   &   &   &   &   &   &   &   &   &   &  &   \\ \hline
            \textbf{QFS-Net}              &   &   &   &   &   &   &   &   &   &   &   &   & X &   &   &   &   &   &   &   &   &   &   &  &   \\ \hline
            \textbf{CMN}                  &   &   &   &   &   &   &   &   &   &   &   &   &   & X &   &   &   &   &   &   &   &   &  &   &   \\ \hline
            \textbf{QDCNN}                &   &   &   &   &   &   &   &   &   &   &   &   &   &   & X &   &   &   &   &   &   &   &   &  &   \\ \hline
            \textbf{QBP}                  &   &   &   &   &   &   &   &   &   &   &   &   &   &   &   & X &   &   &   &   &   &   &   &  &   \\ \hline
            \textbf{ffNN}                  &   &   &   &   &   &   & X &   &   &   &   &   &   &   &   &   & X &   &   &   &   &   &   &   &   \\ \hline
            \textbf{LSTM-QNN}             &   &   &   &   &   &   &   &   &   &   &   &   &   &   &   &   &   & X & X &   &   &   &  X &  X &  \\ \hline
            \textbf{QGAN}                 &   &   &   &   &   &   &   &   &   &   &   &   &   &   &   &   &   &   &   & X & X & X  &  &  &   \\ \hline
    \caption{Relation Between Paper and Used Algorithm}
	\label{tab:Tabla3Columnas}
\end{longtable}
\end{sidewaystable}

\begin{sidewaystable}
\begin{longtable}
		{|p{2,5cm}  p{0,3cm}  p{0,3cm} p{0,3cm} p{0,3cm} p{0,3cm} p{0,3cm} p{0,3cm} p{0,3cm} p{0,3cm} p{0,3cm} p{0,3cm} p{0,3cm} p{0,3cm} p{0,3cm}|} \hline
		   \textbf{Algorithm} & 
              \cite{Gyongyosi2019} & 
              \cite{bausch2020recurrent} & 
		   \cite{CHEN2020105863} &
              \cite{Ullah_2022} &      
              \cite{Alam_2022} &      
              \cite{Du_2022} &     
              \cite{Reya_2022} & 
              \cite{cherrat2022quantum} & 
              \cite{heese2023explaining} &
              \cite{Tilly_2022} &
              \cite{Zoufal_2021} & \cite{Diep2020} &
              \cite{SKAVYSH2023104680} & 
              \cite{KIM2023126643}\\ \hline

            \textbf{RQNN}                  & X & X &   &   &   &   &   &   &   & &  &  &  &  \\ \hline
            \textbf{QREDNN}                &   &   & X &   &   &   &   &   &   & &  &  &   & \\ \hline
            \textbf{QCNN}                  &   &   &   & X & X &   &   &   &   & &  &  &  & X \\ \hline
            \textbf{VQA}                   &   &   &   &   &   & X &   &   &   & &  &   &  &  \\ \hline
            \textbf{QPSO}                  &   &   &   &   &   &   & X &   &   & &  &   &  & \\ \hline
            \textbf{Transformers}          &   &   &   &   &   &   &   & X &   & &  &   &  & \\ \hline
            \textbf{QSVM}                  &   &   &   &   &   &   &   &   & X & &  &  &   & \\ \hline
            \textbf{VQE}                   &   &   &   &   &   &   &   &   &   & X &  &   &  & \\ \hline    
            \textbf{QBM}                   &   &   &   &   &   &   &   &   &  &  & X &  X  &  & \\ \hline    
            \textbf{QMonteCarlo}           &   &   &   &   &   &   &   &   &  &  &  &    & X & \\ \hline               
    \caption{Relation Between Paper and Used Algorithm}
	\label{tab:Tabla3Columnas}
\end{longtable}
\end{sidewaystable}

\subsubsection{Quantum Boltzmann machines (QBM)}
A QBM is defined by a Hamiltonian model of the form $H_{\theta} = \sum^{p - 1}_{i=0}\theta_{i}h_{i}$ where $\theta \in {\rm I\!R}^{p}$ y $h_{i} = \otimes^{n - 1}_{j=0} \sigma_{j,i}$ for $\sigma_{j,i} \in \{ I, X, Y, Z \}$ acting in the $j^{th}$ qubit. Equivalent to classical Boltzmann machines \cite{Ackley1985}, QBMs are typically represented by an Ising model.

The nodes of the network, given by the Pauli terms $\sigma_{j,i}$, are defined concerning certain subsets of qubits. The qubits that determine the output of the QBM are the visible qubits, while the others correspond to the hidden qubits. The probability of measuring a configuration $\upsilon$ of the visible qubits is defined with respect to the  $\Lambda_{\upsilon} = \ket{\upsilon}\bra{\upsilon} \otimes I$ in the quantum Gibbs state \cite{Zoufal_2021,Diep2020}.

\begin{equation}
    \rho^{Gibbs} = \frac{e^{-H_{\sigma}/(k_{B}T)}}{Z}
\end{equation}

with $Z = Tr[e^{-H_{\sigma}/(k_{B}T)}]$, i.e., the probability of measuring $\ket{\upsilon}$ is given by

\begin{equation}
    p^{QBM}_{\upsilon} = Tr[\Lambda_{\upsilon}\rho^{Gibbs}]
\end{equation}

The Gibbs state $\rho^{Gibbs}$ describes the probability of a system being in a certain state as a function of that state's energy and the system's temperature $T$.

\subsubsection{Variable depth quantum circuits (vVQC)}
In \cite{Huang2020}, the authors aim to automatically change the quantum circuit structure according to the score of the qBAS metric \cite{benedetti2019_2}, a complementary metric designed for benchmarking quantum-classical hybrid systems, to solve the problem of the inflexibility of the Variational Quantum Classifier (VQC) model \cite{Schuld2020}. Furthermore, in this paper, the depth of the quantum circuit (the number of quantum circuit layers) is considered a parameter during training.

\begin{figure}[!htbp]
\centering
\includegraphics[width=0.55\textwidth]{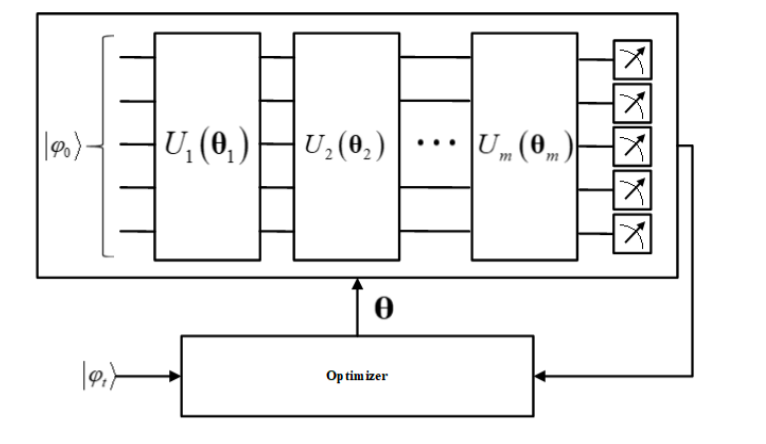}
\caption{VQC Framework \cite{Huang2020}}
\label{fig:1_from_Huang2020}
\end{figure}

The quantum circuit block is taken $U_{l}(\theta_{l})$ as a layer, which is constructed using the circuit block parameterized in [fig:\ref{fig:1_from_Huang2020}]. It is constructed using $e^{ - i\theta\sigma}$ rotation gates, where $\sigma$ is chosen from the set of $\{ X, Y, Z\}$ Pauli gates, which are applied at each qubit. The $\theta$ degree of the rotation is the circuit parameter to be tuned.

\begin{figure}[!htbp]
\centering
\includegraphics[width=0.55\textwidth]{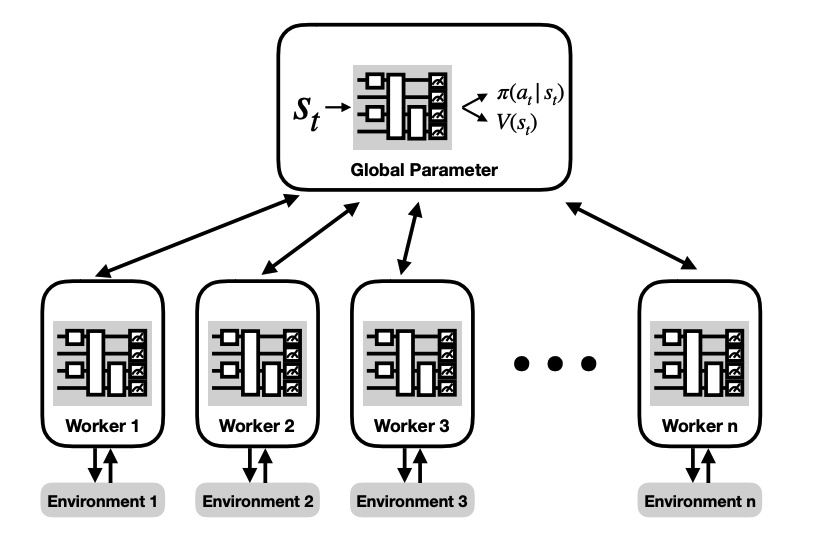}
\caption{Quantum asynchronous advantage actor-critic (A3C) learner \cite{CHEN2023321}}
\label{fig:2023_2_from_CHEN2023321}
\end{figure}

Also, in \cite{CHEN2023321} a framework [fig:\ref{fig:2023_2_from_CHEN2023321}] to train asynchronously these circuits is presented.
The quantum asynchronous advantage actor-critic (QA3C) framework includes a global shared parameters and multiple parallel workers. The results of the gradients of each local agent are stored in the process-specific memories; after that, they are uploaded to the global shared memory, where the global model parameters are updated. To conclude, the updated global model parameters are then immediately downloaded to the local agent that just uploaded its own gradients.

\subsubsection{Hybrid quantum autoencoders (HQA)}
A variation of the Quantum Amplitude Estimation (QAE) algorithm \cite{Brassard2002} called hybrid quantum autoencoder (HQA) is proposed in \cite{Srikumar_2022}. This model incorporates both classical machine learning, in the form of neural networks (ANNs), and quantum machine learning, using quantum neural networks (QNNs) based on parameterized quantum circuits (PQC).

The general design of the model, which is a combination of an encoder that takes a quantum state from Hilbert space $H^{\otimes n}_{2}$ to a subset of real vector space $V$ of dimension $\upsilon = dim(V)$, and a decoder that performs the inverse of that operation.

\begin{figure}[!htbp]
\centering
\includegraphics[width=0.99\textwidth]{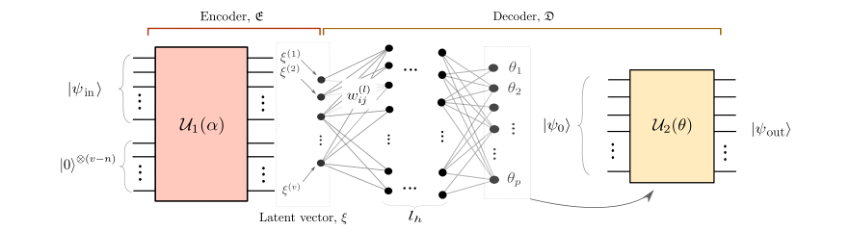}
\caption{Hybrid Quantum Autoencoder (HQA) \cite{Srikumar_2022}}
\label{fig:2_from_Srikumar_2022}
\end{figure}

Although the functional form of the encoder and decoder are defined, the models themselves are not specified. As seen in [fig:\ref{fig:2_from_Srikumar_2022}], the $\epsilon$ encoder is a quantum circuit parameterized by a vector $\alpha$. The circuit receives some $\ket{\psi_{in}}$ state and applies the unitary $U_{1}(\alpha)$ on the combined system of the input state with $(\upsilon - n)$ auxiliary qubits.

\subsubsection{Quantum reservoir computing (QRC)}
Quantum reservoir computing (QRC) uses a dynamic system to execute temporal information processing tasks \cite{suzuki2021natural}. The main objective of this task, called forecasting, is creating a function to transform an input sequence (time series) into a target output sequence for time-series prediction and pattern classification.

In the QRC framework, the dynamical system is given by
\begin{equation}
    \rho_{t} = T_{u_{t}}(\rho_{t} - 1),
\end{equation}

where $\rho_{t}$ is the density operator representing a state at time t, and $T_{u_{t}}$ is an input-dependent, fully positive, trace-preserving map (CPTP) describing the time evolution of the QR system. The considered CPTP map is given by

\begin{equation}
    \rho_{t} = e^{- iH\tau} (\rho_{input} \otimes Tr_{input}(\rho_{t - 1})) e^{iH\tau},
\end{equation}

where $\rho_{input} = \ket{\psi_{u_{t}}} \braket{\psi_{u_{t}}|  with  |\psi_{u_{t}}} = \sqrt{1 - u_{t}} \ket{0} \bra{0} + \sqrt{u_{t}} \ket{1} \bra{1}$. An auxiliary qubit takes the input value $u_{t} \in [0, 1]$, and then the whole QR system is transformed in time by the input-independent unitary operator $e^{-iH\tau}$.

\subsubsection{Quantum multiclass classifier (QMCC)}
\begin{figure}[!htbp]
\centering
\includegraphics[width=0.7\textwidth]{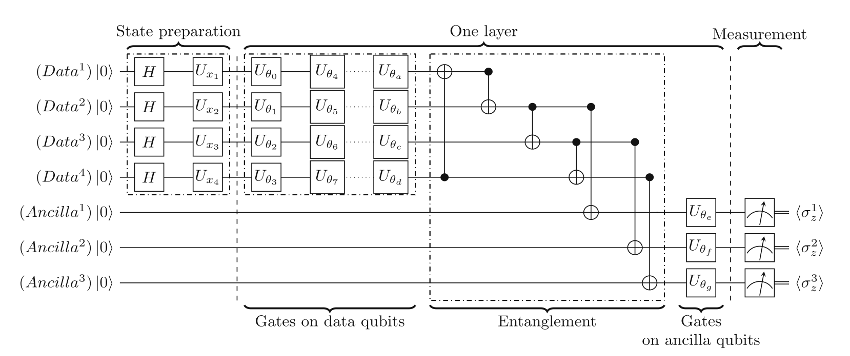}
\caption{Quantum Variational Circuit Model \cite{Chalumuri2021}}
\label{fig:3_from_Chalumuri2021}
\end{figure}

In \cite{Chalumuri2021}, a variational circuit designed to classify the dataset with four features is proposed [fig:\ref{fig:3_from_Chalumuri2021}]. The operations performed in the variational circuit are as follows. The four qubits of the circuit are initialized in the $\ket{0}$ state. Next, the Hadamard gate is applied individually to place the qubits in a superposition of $\ket{0}$ and $\ket{1}$. Secondly, a unitary operation is applied to each qubit with a unitary square matrix designed for state preparation. In this way, the classical data (bits) are encoded into qubits.

\begin{figure}[!htbp]
\centering
\includegraphics[width=0.5\textwidth]{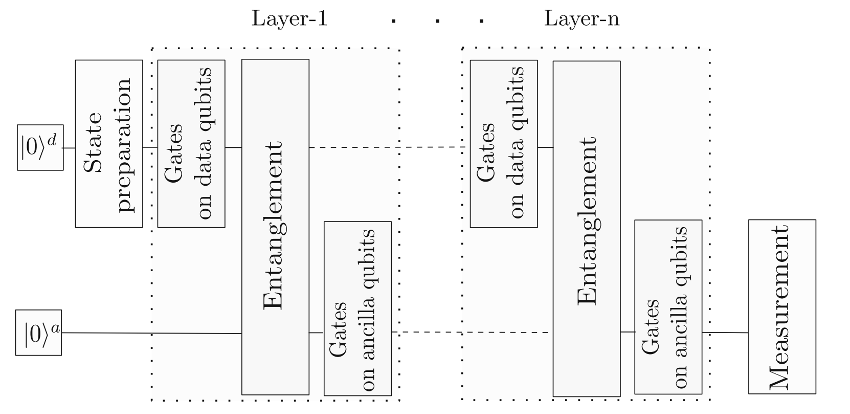}
\caption{Architecture of the Quantum Variational Circuit \cite{Chalumuri2021}}
\label{fig:4_from_Chalumuri2021}
\end{figure}

After state preparation, the variational circuit is designed using multiple layers of interleaved rotational gates in data qubits and auxiliary qubits with adjustable parameters using an optimization technique. The architecture of the variational circuit model and the original implementation of the circuit, which consists of seven layers, is depicted in [fig:\ref{fig:4_from_Chalumuri2021}]. Finally, auxiliary qubits are measured, and resulting qubits are processed to obtain the class label.

\subsubsection{Support vector machines with a quantum kernel estimator (QSVM-Kernel)}
The quantum support vector machine QSVM and the quantum kernel estimator (QSVM-Kernel) \cite{wu2021application} exploits the quantum state space as a feature space to compute kernel inputs efficiently. This algorithm maps the classical data $\vec{x}$ nonlinearly to a quantum state of N qubits by applying a quantum circuit $\Gamma_{\Phi(\vec{x})}$ to the initial state $\ket{0^{\otimes N}}$:

\begin{equation}
    \ket{\Phi(\vec{x})} = \Gamma_{\Phi(\vec{x})}\ket{0^{\otimes N}}
\end{equation}

\begin{figure}[!htbp]
\centering
\includegraphics[width=0.55\textwidth]{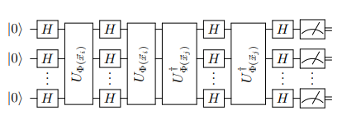}
\caption{Quantum Circuit of Support Vector Machine \cite{wu2021application}}
\label{fig:5_from_wu2021application}
\end{figure}

The quantum circuit $\Gamma_{\Phi(\vec{x})}$ gives rise to a $2^{N}$ dimensional feature space (N is the number of qubits), which is hard to estimate classically. This circuit consists of two repeated layers.

\begin{equation}
     \Gamma_{\Phi (\vec{x})} = U_{\Phi (\vec{x})} H^{\otimes N} U_{\Phi (\vec{x})} H^{\otimes N}
\end{equation}

where H is a Hadamard gate and $U_{Phi(\vec{x})}$ is a unitary operator that encodes the classical input data [fig:\ref{fig:5_from_wu2021application}].

The kernel entries are evaluated for the training data and used to find a separation hyperplane in the training phase. Then, in the test phase, the kernel inputs are evaluated with new data $\vec{x}$ and with the training data, which are used to classify the new data $\vec{x}$ according to the separation hyperplane.

The kernel inputs are evaluated on quantum computers, while the optimization of the separation hyperplane and the classification of the data are performed on classical computers, as in a classical SVM.

Regarding quantum kernels, in \cite{Huang_2021}, the authors introduced projected quantum kernels and outperformed all tested classical models in prediction error. First, they use principal component analysis \cite{Jolliffe_2004} to transform each image into a $n$-dimensional vector. Second, they embed this vector into a quantum circuit using three techniques, a separable rotation circuit \cite{Schuld_2020, Schuld_2019, Skolik_2021}, an IQP circuit \cite{Havl_ek_2019}, and a Hamiltonian evolution circuit.

\subsubsection{Quantum neural networks QNN}

\begin{figure}[!htbp]
\centering
\includegraphics[width=0.4\textwidth]{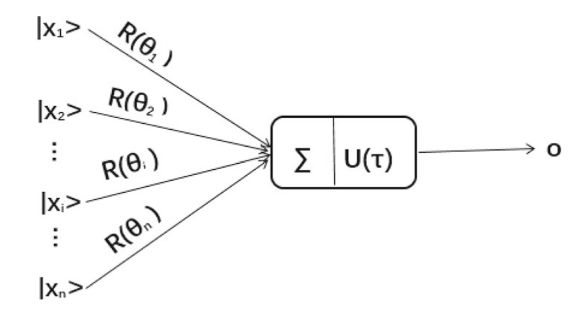}
\caption{Quantum Neuron Model \cite{Chen2020b}}
\label{fig:6_from_Chen2020b}
\end{figure}

Quantum neural networks are composed by quantum neurons, quantum neuron model is described in \cite{Chen2020b}. The input of this model is the quantum state $\ket{x_{i}}$, and the output is denoted as real number $o$ [fig:\ref{fig:6_from_Chen2020b}]. The value of the quantum weight $R(\theta_{i} )$ is the quantum rotation gate $R(\theta) = 
\begin{bmatrix}
cos\theta & -sin\theta\\
sin\theta & cos\theta
\end{bmatrix}$, Controlled-NOT gate $U(\tau) = C(f(\tau))$, where $f$ is the sigmoid function, and the function $C(x)$ is $C(k)\ket{\phi} = 
\begin{bmatrix}
cos(\frac{\pi}{2}m - \alpha) \\
sin(\frac{\pi}{2}m - \alpha)
\end{bmatrix}$ [fig:\ref{fig:6_from_Chen2020b}].

Once the quantum neuron model is introduced, the model of quantum neural networks is presented, \cite{Tacchino_2021,Bausch2018} where classical input and weight vectors of size $m$ can be encoded in quantum hardware using only $N = log_{2} m$ qubits. Relative quantum phases (i.e., $\pm 1$) in superpositions of base states can be used to encode collections of classical data.

The network training can be done in two ways, either by a global variational training, where the quantum artificial neuron is obtained by inverting a non-trivial quantum state $\ket{\psi_{w}}$ back to the base state $\ket{1}^{\otimes N}$, or by local variational training, in which global transformation $V(\vec{\theta})$ is divided into successive layers $V_{j}(\vec{\theta_{j}})$ of decreasing complexity and size. Each layer is trained separately, according to a cost function applied to obtain the desired final state.

Transfer learning for QNN is proposed in \cite{Mari_2020} and in \cite{KIM2023126643}. The authors use an effective strategy for small QCNNs in the noisy intermediate-scale quantum era. Allowing QCNN to solve complex classification problems by utilizing a pre-trained classical convolutional neural network.

\subsubsection{Hybrid k-neighbours-nearby model (HKNN)}
\begin{figure}[!htbp]
\centering
\includegraphics[width=0.3\textwidth]{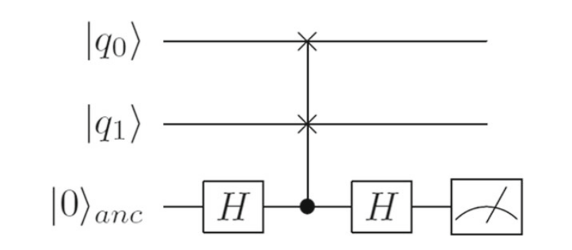}
\caption{Test Swap Circuit\cite{LaBorde2020}}
\label{fig:7_from_LaBorde2020}
\end{figure}

In quantum states, the overlap between two states acts as a similarity measure analogous to the Euclidean distance, and this overlap is found through a simple circuit known as a swap-test, as shown in [fig:\ref{fig:7_from_LaBorde2020}]. This circuit can be used to evaluate the distances between classical vectors in KNN algorithms \cite{LaBorde2020}. 

\begin{figure}[!htbp]
\centering
\includegraphics[width=0.4\textwidth]{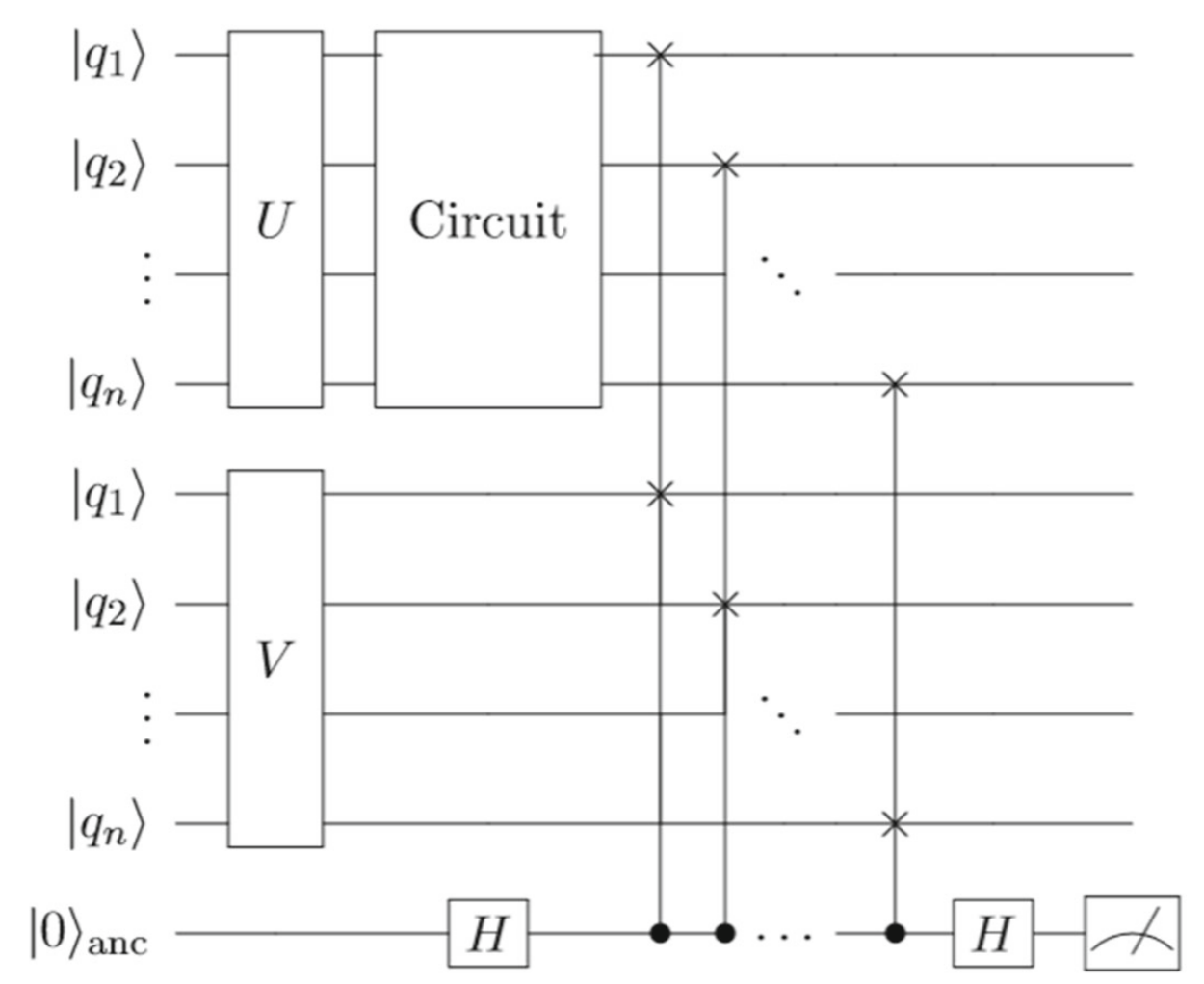}
\caption{QKNN Circuit \cite{LaBorde2020}}
\label{fig:23_from_LaBorde2020}
\end{figure}

This circuit takes input data that is in the state $\ket{0}$ or $\ket{1}$, of the form $\ket{q_{1}}\ket{q_{2}} ... \ket{q_{n}}$. Then, successive SWAP-controlled gates or Fredkin gates controlled with an auxiliary qubit are applied. The reference state is sent to unitary qubit $V$ [fig:\ref{fig:23_from_LaBorde2020}], generating a test circuit instead of the reference circuit.

Comparing the qubit of the reference state to its corresponding one sent to the test circuit, if the output state of the test circuit $\ket{\psi}$ is the same as the reference state $\ket{\phi}$, then the probability of measuring the auxiliary bit to be in the state $\ket{0}$ is one, however, if the states are slightly different, the probability of measuring the auxiliary bit in the zero state is determined by :

\begin{equation}
    P(\ket{0}_{anc}) = \frac{1}{2} + \frac{1}{2}|\braket{\psi|\phi}|^{2},
\end{equation}

The generalization for multiple qubits is given by:

\begin{equation}
    P(\ket{0}_{anc}) = \frac{1}{2^{n}} + \frac{1}{2^{n}}\sum^{n}_{i=1}|\braket{\psi_{i}|\phi_{i}}|^{2},
\end{equation}

where the respective $i$-th qubits of the two states are summed, and $n$ is the number of qubits introduced into the test circuit. This probability is the overlap between the two states. Using $d$ different pairs of input and reference states, a classical $d$-dimensional vector \textbf{S} is constructed containing the measured probability for each set of states such that:

\begin{equation}
    \textbf{S} = [P_{1}, P_{2},...,P_{d}]
\end{equation}

\subsubsection{Orthogonal neural networks}

Orthogonal neural networks have recently been introduced as a new neural network type that enforces orthogonality in weight matrices \cite{jia2019orthogonal, wang2020orthogonal}. Several classical gradient descent methods have been proposed to preserve orthogonality while updating weight matrices, but these techniques suffer from long runtimes and provide only approximate orthogonality.

The idea of orthogonal neural networks (OrthoNNs) is to add a constraint to the weight matrices corresponding to the layers of a neural network, with algorithms such as the Stiefel manifold \cite{jia2019orthogonal}. However, the main difficulty with OrthoNNs is preserving the orthogonality of the matrices while updating during gradient descent.

A new type of neural network layer called Pyramidal Circuit is introduced in \cite{kerenidis2021classical}, which implements orthogonal matrix multiplication. It allows gradient descent with perfect orthogonality with the same execution time as a standard layer.

The proposed quantum circuit implements a fully connected neural network layer with an orthogonal weight matrix, using only one type of quantum gate, the Reconfigurable Beam Splitter (RBS) gate. This two-qubit gate is parameterizable with an angle $\theta \in [0, 2\pi]$. Its matrix representation is given by:

\begin{equation}
RBS(\theta) = 
\begin{pmatrix}
1 & 0 & 0 & 0\\
0 & cos\theta & sin\theta & 0\\
0 & -sin\theta & cos\theta & 0\\
0 & 0 & 0 & 1
\end{pmatrix}    
RBS(\theta) : \left\lbrace\begin{array}{c} \ket{01} \mapsto cos\theta \ket{01} - sin\theta \ket{10} \\ \ket{10} \mapsto sin\theta \ket{01} + cos\theta \ket{10} \end{array}\right.
\end{equation}

\begin{figure}[!htbp]
\centering
\includegraphics[width=0.9\textwidth]{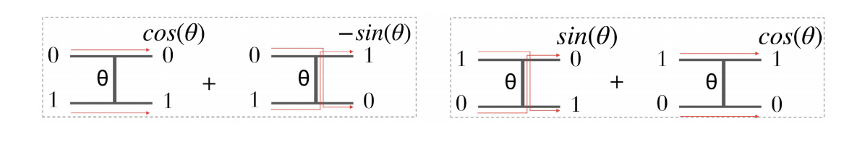}
\caption{Quantum Mapping in Two Qubits \cite{kerenidis2021classical}}
\label{fig:8_from_kerenidis2021classical}
\end{figure}

Starting from two qubits, one in the $\ket{0}$ state and the other in the $\ket{1}$ state, the qubits may or may not be swapped in superposition. The $\ket{0}$ qubit remains in its thread with the amplitude $cos$ or, conversely, is swapped with the other qubit if the amplitude is $- sin\theta$ while the new thread is above $(\ket{01} \mapsto \ket{10})$, or $+ sin\theta$ if the new thread is below $(\ket{01} \mapsto \ket{10})$. In the other two cases $(\ket{00} \mapsto \ket{11})$ the gate $RBS$ acts as an identity [fig:\ref{fig:8_from_kerenidis2021classical}].

This gate can be implemented either as a single gate known as the $FSIM$, or using four Hadamard gates, two $R_{y}$ rotation gates, and two two-bit $CZ$ gates [fig:\ref{fig:9_from_kerenidis2021classical}]:

\begin{figure}[!htbp]
\centering
\includegraphics[width=0.3\textwidth]{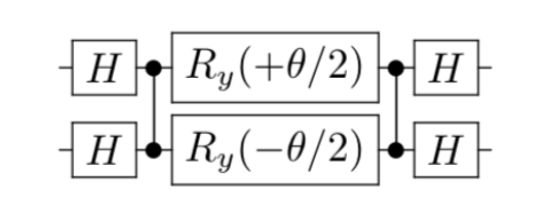}
\caption{Decomposition of the RBS Door \cite{kerenidis2021classical}}
\label{fig:9_from_kerenidis2021classical}
\end{figure}

\subsubsection{Quantum fully self-supervised neural networks (QFS-Net)}

In \cite{Konar2021}, a quantum neural network model based on qudits is used. A qudit is a multilevel quantum system (D $>$ 2) represented by D basis states: $\ket{0}, \ket{1}, \ket{2}, ..., \ket{D - 1}$, in which the superposition of these basis states would be represented by:

\begin{equation}
    \ket{\psi} = a_{0}\ket{0} + a_{1}\ket{1} + a_{2}\ket{2} + ... + a_{D - 1}\ket{D - 1} =  \begin{bmatrix}
                                                a_{0}\theta \\
                                                a_{1}\theta \\
                                                ...\\
                                                a_{D - 1}
                                            \end{bmatrix}
\end{equation}

with the normalisation criterion $|a_{0}|^{2} + |a_{1}|^{2} + ... + |a_{D - 1}|^{2} = 1$.

In this type of qubit-based quantum neural network model, the classical inputs to the network become D-dimensional given by $x_{k}$. A sigmoid function is applied to the quantum states $[0,(2\pi/D)]$ as long as the $k$-th input is the function $f_{QNNM}(x_{k})$, yielding a binary classical result [0, 1].

\begin{equation}
    f_{QNNM}(x_{k}) = \frac{1}{1 + e^{-x_{k}}}
\end{equation}

Then, map that result to the amplitude in the $k$-th ground state

\begin{equation}
    \ket{a_{k}} = \left(\frac{2\pi}{D}f_{QNNM}(x_{k})\right)
\end{equation}

\begin{figure}[!htbp]
\centering
\includegraphics[width=1.00\textwidth]{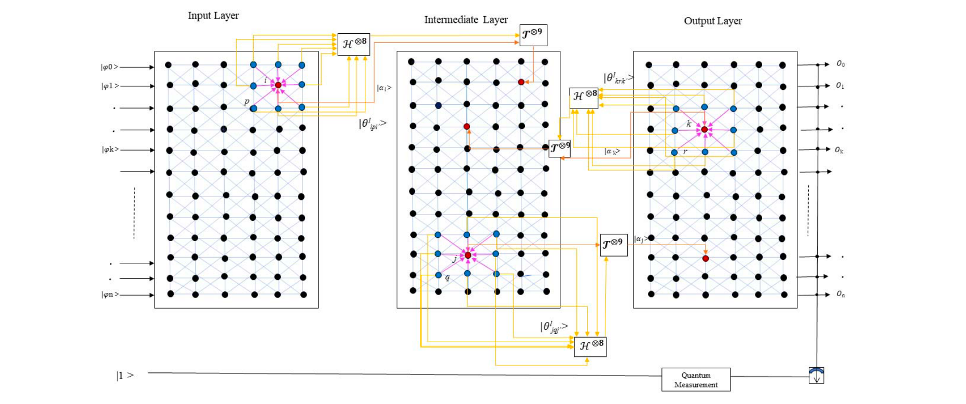}
\caption{QFS-Net \cite{Konar2021}}
\label{fig:10_from_Konar2021}
\end{figure}

An example of such a network is the QFS-Net architecture \cite{Konar2021}, where the information processing unit is the qutrit. The qutrit neurons in each layer are obtained using a transformation gate $T$, and the interconnection weights are mapped onto the qutrits using Hadamard gates ($H$). The rotation angle (pink arrow in [fig:\ref{fig:10_from_Konar2021}]) is adjusted by obtaining the relative difference of the information between each candidate qutrit neuron and the neighboring qutrit neuron of the same layer used in the rotation gate to update the interconnections between layers. A new sigmoid activation function guides the self-propagation and counter-propagation mechanisms of the QFS network.

\subsubsection{CNOT neural networks (CMN)}

Although quantum neural network models exist, their computational complexity may require specific unitary transformations to simulate the neuron's activation function or a large number of auxiliary qubits to perform the desired transformations. To solve some of these problems, a quantum neural network model called the CNOT measured network (CMN) is presented in \cite{Lukac2018}.

The CMN uses only CNOT quantum gates and the measurement operator, which are very simple to implement in any quantum computing technology. The CMN can, using only these two simple operators, accommodate universal operators such as AND and OR while maintaining an optimized learning rate and a constant number of auxiliary qubits.

The main objective of the proposed model allows reducing the computational complexity required and minimize the number of different computational elements. In order to achieve this objective, only discrete Boolean functions are designed, and models with a larger number of layers with simple transfer functions are chosen.

The phases of the network creation process would be the initialization procedure for an arbitrary quantum state, the implementation of the CNOT gate, and the addition of the measurement operator. 


\begin{figure}[!htbp]
\centering
\includegraphics[width=0.6\textwidth]{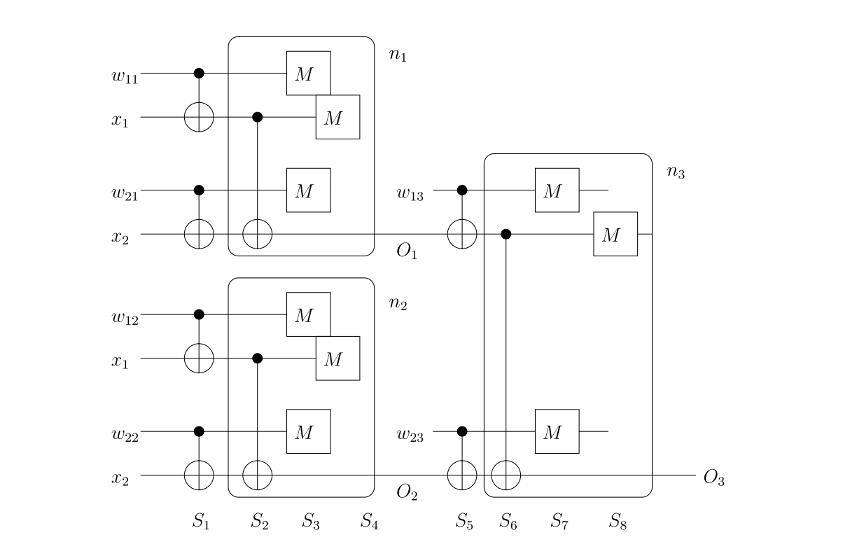}
\caption{CNOT Quantum Neural Network\cite{Lukac2018}}
\label{fig:12_from_Lukac2018}
\end{figure}

In the information processing stage, two types of qubits are distinguished, data qubits and weight qubits. Data qubits $x_{1}$  and $x_{2}$ are initialised to a state corresponding to the value of the input variables. For example, $\ket{x_{1}} = \ket{0}$. The weights are initialised with a random value as $w_{i} = \frac{1}{\sqrt{3}}\ket{0} + \frac{\sqrt{2}}{\sqrt{3}}\ket{1}$ [fig:\ref{fig:12_from_Lukac2018}].

The data qubits are then interleaved with the weight qubits using the $CNOT$ quantum gate as follows:

\begin{equation}
  \begin{aligned}
    \ket{\phi} &= CNOT\ket{w_{ji} \otimes x_{i}} \\
     &= \alpha_{w_{ji}} \alpha_{x_{i}} \ket{0}_{w_{ji}} \otimes \ket{0}_{x_{i}} \pm \beta_{w_{ji}} \beta_{x_{i}} \ket{1}_{w_{ji}} \otimes \ket{1}_{x_{i}}
  \end{aligned}
\end{equation}

The output function can be expressed using the notation of the density operator as a reduced density operator, described by

\begin{equation}
    \rho = \sum_{i}p_{i}\ket{\psi_{i}}\bra{\psi_{i}}
\end{equation}

And the indices $i$ the measurement of the qubits $\ket{x_{0}w_{0}...x_{n}w_{n}}$ can be represented by the trace of the density matrix:

\begin{equation}
  \begin{aligned}
    \rho^{o} &= \sum_{i}p_{i}tr(\ket{\psi_{i}}\bra{\psi_{i}}) \otimes  \sum_{i}p_{j}\ket{o_{j}}\bra{o_{j}} \\
    \rho^{o} &= \sum_{i}p_{j}\ket{o_{j}}\bra{o_{j}}
  \end{aligned}
\end{equation}

\subsubsection{Quantum convolutional deep convolutional neural networks (QDCNN)}

Another type of algorithm is the quantum adaptation of convolutional networks. For example, in \cite{Li2020b}, we can see an implementation of a deep convolutional quantum convolutional neural network (QDCNN).

The first step of QDCNN is preparing the classical input image in a quantum state. Given a 
$2^{n}$ x $2^{n}$ classical grey input image $I$, it can be described as a matrix:

\begin{equation}
   \begin{bmatrix}
    c_{0,0} & c_{0,1} & \ldots & c_{0,N-1}\\
    c_{1,0} & c_{1,1} & \ldots & c_{0,N-1}\\
    \vdots & \vdots & \vdots & \vdots \\
    c_{N-1,0} & c_{N-1,1} & \ldots & c_{N-1,N-1}
   \end{bmatrix}
\end{equation}

where $N = 2^{n}$ and the elements $c_{x,y} \in [0,255]$ represents the value of the corresponding pixel at position $(x,y) \in [0,N-1]^{\otimes 2}$

Analogous to the classical situation, the QDCNN architecture consists of several successive quantum convolutional layers and a quantum classifier layer.

First, all qubits in the quantum register are initialized as $\ket{0}$ state. Afterward, Hadamard gates are applied to the $regL$ quantum register and the QRAM algorithm to the $regC$ and $regK$ quantum registers to prepare the input image and the convolutional kernel, respectively. 

\begin{figure}[!htbp]
\centering
\includegraphics[width=0.65\textwidth]{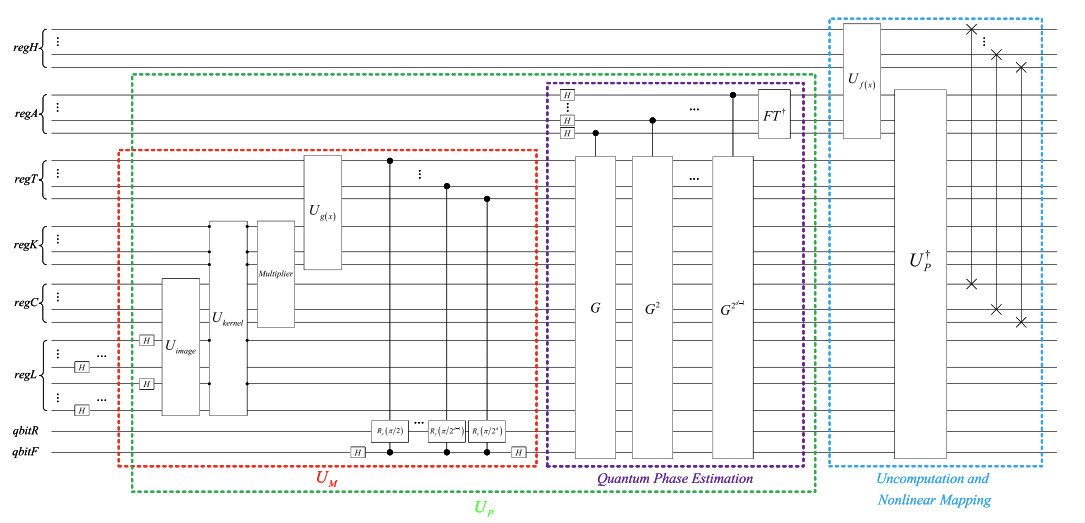}
\caption{QDCNN-Net Architecture \cite{Li2020b}}
\label{fig:21_from_Li2020b}
\end{figure}

Quantum multiplication is applied to the $regC$ and $regK$ registers. A bifurcation qubit $qbitF$ and a rotation qubit $qbitR$ are added [fig:\ref{fig:21_from_Li2020b}].

\begin{equation}
   U_{QRAM} = (\{d_{j}\})\ket{j}\ket{0} \mapsto \ket{j}\ket{d_{j}}
\end{equation}

\begin{figure}[!htbp]
\centering
\includegraphics[width=0.4\textwidth]{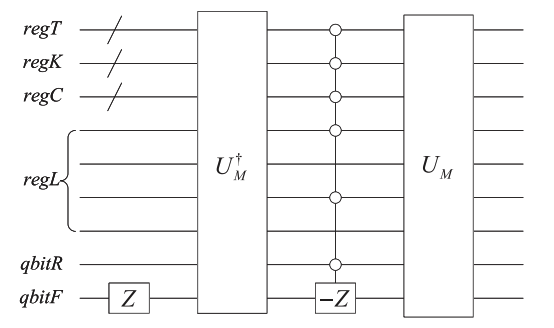}
\caption{QDCNN-Net G Oracle \cite{Li2020b}}
\label{fig:22_from_Li2020b}
\end{figure}

The QPE algorithm of the G oracle \cite{Li2020b}[fig:\ref{fig:22_from_Li2020b}] is implemented on the $regA$ obtaining an estimate of the quantum amplitude (QAE). Finally, an additional quantum register is introduced to make the linear quantum theory compatible with the nonlinear DCNN model, and a nonlinear modulus $U_{f(x)}$ is used.

\subsubsection{Quantum backpropagation neural networks (QBP)}

\begin{figure}[!htbp]
\centering
\includegraphics[width=0.6\textwidth]{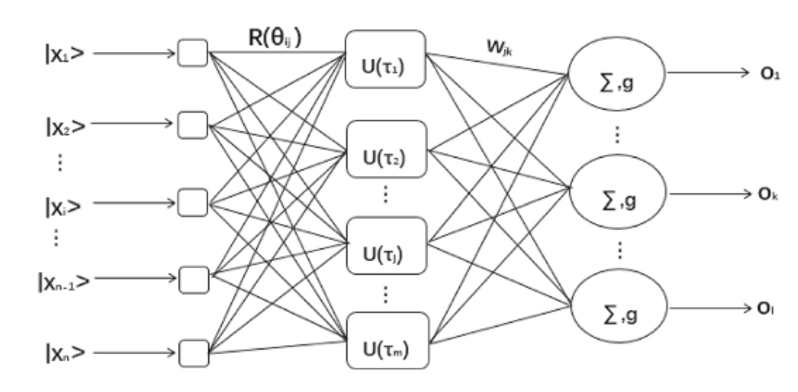}
\caption{Quantum Backpropagation Neural Network \cite{Chen2020b}}
\label{fig:13_from_Chen2020b}
\end{figure}

Based on the quantum neuron model and the structure of the classical back-propagation (BP) neural network, the QBP feedforward neural network [fig:\ref{fig:13_from_Chen2020b}] is constructed in \cite{Chen2020b}.

In this model, the input and hidden neurons are quantum, and the output neurons are classical. The input vector is $X = [\ket{x_{1}} \ket{x_{1}} \ket{x_{i}} \ldots \ket{x_{n}}]^{T}$, the final output vector of the network is $O = [o_{1} o_{2} \ldots o_{k} \ldots o_{l}]^{T}$ and the weight matrix $V = (v_{i,j})_{nm}$, where $v_{ij}$ is the weight between the $i$-th input node and the $j$-th hidden node.

Since the training samples are always real numbers, a specific method for the transfer of real numbers to quantum states is defined. The vector of real samples is $X = [x_{1} x_{2} \ldots x_{i} \ldots x_{n}]^{T}$, thus we can obtain the corresponding input quantum vector $[\ket{x_{1}} \ket{x_{1}} \ket{x_{i}} \ldots \ket{x_{n}} ]^{T}$, where

\begin{equation}
   X_{i} = 
\begin{bmatrix}
cos(2\pi) \cdot sigmoid(x_{i})\\
sin(2\pi) \cdot sigmoid(x_{i})
\end{bmatrix}
\end{equation}

To avoid the local optimization problem in the gradient descent calculation, a genetic algorithm (GA) is used. In contrast to the GA, the QGA \cite{Nicolay2018} can maintain the diversity of the population due to the quantum superposition state, simplifies the calculation, and reduces the operation steps by using the quantum rotational gate. By combining the QGA algorithm and BP algorithm, the training error of the QBP neural network can be minimized, and we can prevent the training from falling into a local optimization situation.

In this model, each chromosome is encoded by qubits in the following way

\begin{equation}
    q^{t}_{j} = 
   \begin{bmatrix}
    \alpha^{t}_{11} & \ldots & \alpha^{t}_{1k} & 
    \alpha^{t}_{21} & \ldots & \alpha^{t}_{2k} & 
    \ldots &
    \alpha^{t}_{m1} & \ldots & \alpha^{t}_{mk}\\
    \beta^{t}_{11} & \ldots & \beta^{t}_{1k} & 
    \beta^{t}_{21} & \ldots & \beta^{t}_{2k} & 
    \ldots &
    \beta^{t}_{m1} & \ldots & \beta^{t}_{mk}
   \end{bmatrix}
\end{equation}
where $q^{t}_{j}$ is the $j$-th chromosome of the $t$-th generation, $k$ represents the number of qubits encoded by each gene, m is the number of genes in the chromosome.

\subsubsection{Feed foward neural networks (ffNN)}

The architecture of quantum ffNNs is defined in \cite{Tacchino2020, Tacchino_2021}. An ffNN is essentially composed of a set of individual ${n_{i}}$ nodes, or artificial neurons, arranged in a sequence of successive ${L_{j}}$ layers. Information flows through the network from the input layer to the output layer, traveling through connections between neurons.

Artificial neurons combine the input $(\overrightarrow{i})$ and the weight $(\overrightarrow{i})$.
$(\overrightarrow{w})$, providing an activation that depends on their scalar product $\overrightarrow{i} \cdot \overrightarrow{w}$. 

Assuming that the quantum register of $N$-qubits is initially in the rest configuration, $\ket{0}^{\otimes N}$, the quantum state of the input vector with a unitary operation is prepared $U_{i}$ such that $|\psi_{i}> = U_{i}\ket{0}^{\otimes N}$. Next, the weight factors of the vector are applied $\overrightarrow{w}$ on the input state by implementing another unitary transformation, $U_{w}$, subject to the constraint $\ket{1}^{\otimes N} = U_{w}\ket{\psi_{w}}$. 

When several copies of the quantum register run in parallel, these, and the result of measurements made on them, can be used to feed information about the processing of the input weight to a successive layer.

\begin{figure}[!htbp]
\centering
\includegraphics[width=0.5\textwidth]{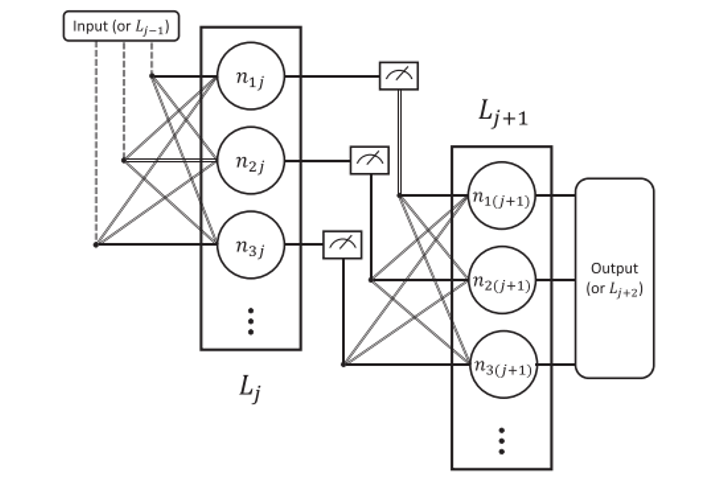}
\caption{Feed Foward Neural Network (ffNN) \cite{Tacchino2020}}
\label{fig:14_from_Tacchino2020}
\end{figure}

An abstract representation of the proposed architecture is shown in [fig:\ref{fig:14_from_Tacchino2020}].

In \cite{P_rez_Salinas_2020}, another implementation of ffNN is proposed, where the data is introduced in each processing unit joined with the output of the previous processing unit. The data is uploaded multiple times along the circuit by using one-qubit rotations. These rotations are optimized using a classical minimization algorithm.

\subsubsection{Long short-term memory neural networks LSTM-QNN}

\begin{figure}[!htbp]
\centering
\includegraphics[width=0.55\textwidth]{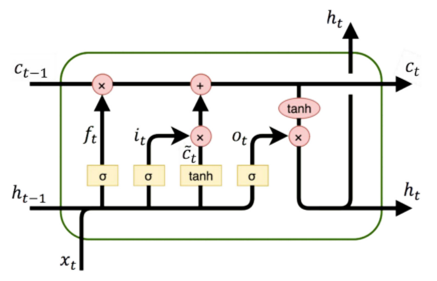}
\caption{LSTM Cell \cite{Ceschini_2022}}
\label{fig:23_from_Ceschini2022}
\end{figure}

In classical deep learning, Long short-term memory (LSTM) networks are used to detect both short and long-term dependencies in entire sequences of data. One of the implementations of the LSTM cell [fig:\ref{fig:23_from_Ceschini2022}] has three inputs; the first one is represented by $x_{t}$, where t is the time index of data samples that is also associated with the specific cell of the unrolled chain, the second input $h_{t-1}$ is the ‘hidden’ state computed at the previous cell (time step) and the third input $c_{t-1}$ is the previous ‘cell’ state. In addition, the cell has three control state cell gates: an input gate (i), a forget gate (f), and an output gate (o). Also, each gate has the sigmoid $\sigma$ or the hyperbolic tangent $\tanh$ activation function \cite{Ceschini_2022}.

\begin{figure}[!htbp]
\centering
\includegraphics[width=0.8\textwidth]{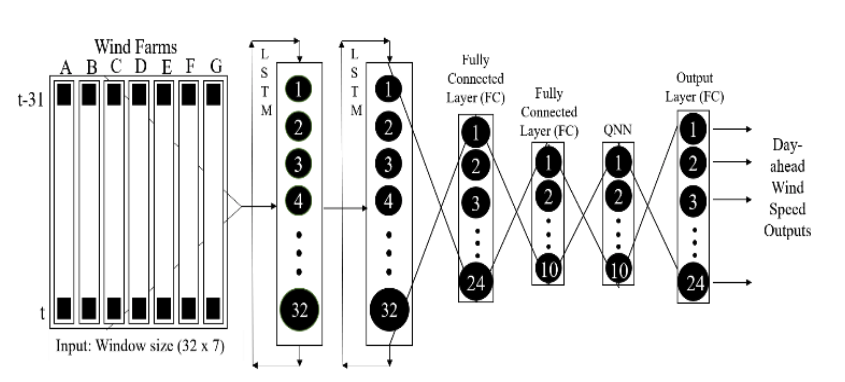}
\caption{LSTM Architecture \cite{Hong2023c}}
\label{fig:2023_1_from_Hong2023c}
\end{figure}

A hybrid architecture is proposed in [fig:\ref{fig:2023_1_from_Hong2023c}] \cite{Hong2023c}. It consists of two layers with 32 cells, two fully connected layers, the second one of 10 neurons, to connect with the QNN. The QNN layer is composed using the IQP Ansatz \cite{Havl_ek_2019} and StronglyEntanglingLayers \cite{Schuld2020}, adding a final output classical layer.

due to the limitation of a 10 qubits layer with 32 cells, it must be followed by a classical layer, which is another dense layer in this example, containing ten neurons, before it can be connected to the QNN with ten qubits. Then, the final predictions may be obtained by adding a fully connected layer after the QNN.

In \cite{Wang2021}, a metalearning quantum approximate optimization algorithm (MetaQAOA) is proposed for the MaxCut problem \cite{Festa2002}. Specifically, a quantum neural network (QNN) in the form of a parameterized quantum circuit and a classical short-term memory (LSTM) neural network is constructed as an optimizer, which can quickly help the QNN \cite{Goncalves2017} to find the approximate optimal parameters of the QAOA. 

The main steps of this algorithm are as follows. A uniformly distributed quantum superposition state is prepared as the initial state $\ket{\psi_{0}}$:

\begin{equation}
    \ket{\psi_{0}} = \ket{+_{1}} \ket{+_{2}} \ldots \ket{+_{n}},
\end{equation}

where $n$ represents the number of nodes in the graph G for the MaxCut problem and $\ket{+} = (1/\sqrt{2})(\ket{0}\ket{1})$ is an overlapping state.

Next, the Hamiltonian operator $\hat{H}_{M}$ and the cost Hamiltonian operator $\hat{H}_{M}$ are selected
 $\hat{H}_{C_{1}}$:

\begin{equation}
    \hat{H}_{M} = \sum_{j \in n}\hat{X}_{j}
\end{equation}
\begin{equation}
    \hat{H}_{C_{1}} = \sum_{\{j,k\} \in E} \frac{1}{2} (\hat{I} - \hat{Z_{j}}\hat{Z_{k}}),
\end{equation}

where $\hat{H}_{M}$ is the Pauli sum $\hat{X}$ in each qubit, $\hat{H}_{C1}$ is diagonal in the computational basis, $I, X,Z$ represent the Pauli operator $I, X,Z$, respectively and $j, k$ are the two nodes connected by the edge of the graph $G$. 

Thirdly, the unitary operator $\hat{U}(\eta , \gamma)$ applied to this initial state given by

\begin{equation}
    \hat{U}(\eta , \gamma) = \prod^{P}_{q=1}e^{-i\eta_{q}\hat{H}_{M}}e^{-i\gamma_{q}\hat{H}_{C_{2}}}
\end{equation}

where $P$ is the number of times the unitary transformation is executed and the parameters $\{\eta , \gamma)\}$ defined by:

\begin{equation}
  \begin{aligned}
    \overrightarrow{\eta} &= (\eta_{1}, \eta_{2}, \ldots, \eta_{p})^{T},\\
    \overrightarrow{\gamma} &= (\gamma_{1}, \gamma_{2}, \ldots, \gamma_{p})^{T}.
  \end{aligned}
\end{equation}

\begin{figure}[!htbp]
\centering
\includegraphics[width=0.6\textwidth]{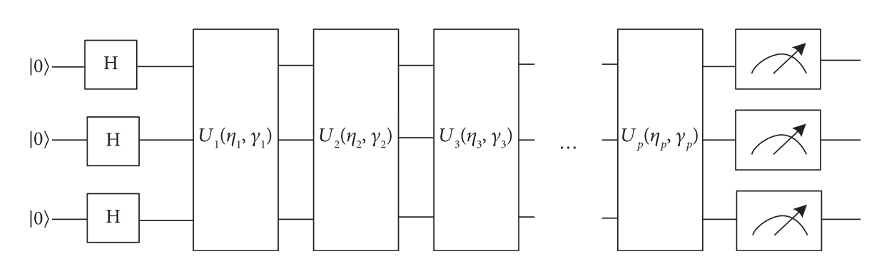}
\caption{LSTM-QNN Parameterised Circuit \cite{Wang2021}}
\label{fig:15_from_Wang2021}
\end{figure}

The resulting parameterized circuit is shown at [fig:\ref{fig:15_from_Wang2021}].

\subsubsection{Quantum generative adversarial networks (QGAN)}

\begin{figure}[!htbp]
\centering
\includegraphics[width=0.6\textwidth]{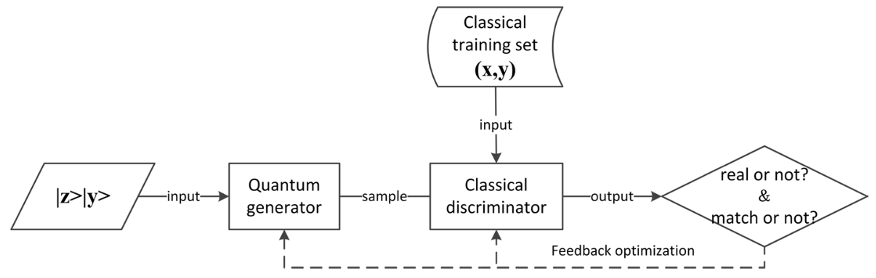}
\caption{Conditional Adversarial Neural Networks \cite{Liu2021}}
\label{fig:16_from_Liu2021}
\end{figure}

In these papers \cite{Situ2020a,Liu2021,Anand2021}, a quantum version of the Generative Adversarial Network (GAN) \cite{Goodfellow2014} is proposed. The input data to the discriminator are the sampled data pairs from the classical training set, where $y$ represents the conditional variable. The generator obtains the conditional variables and the probability distribution of various samples in the training set through $\ket{y}$. Subsequently, the classical discriminator is trained with the training set data and the samples from the quantum generator, whose output feeds back to the quantum generator [fig:\ref{fig:16_from_Liu2021}]. 

At the beginning of the training, all parameters of the quantum circuit and the binary classification neural network are given random initial values. The generator and discriminator parameters are alternately optimized during the training process. The discriminator simultaneously judges the training data and the data sampled from the quantum generator. 

The generator's ability to match the actual distribution is improved by repeating the circuit in the quantum computing device. Optimization of the generator and discriminator parameters must be performed iteratively until the generator can reconstruct the state distribution of the training set.

\begin{equation}
    \ket{y} = \sum^{m}_{j=1}\frac{1}{\alpha_{j}}\ket{y_{j}},
\end{equation}

Regarding the quantum generator circuit design, a PQC is generated by implementing single-qubit gates that are used to perform the rotation of the qubits, as well as multi-qubit gates mainly perform the entanglement between the qubits.

\begin{figure}[!htbp]
\centering
\includegraphics[width=0.6\textwidth]{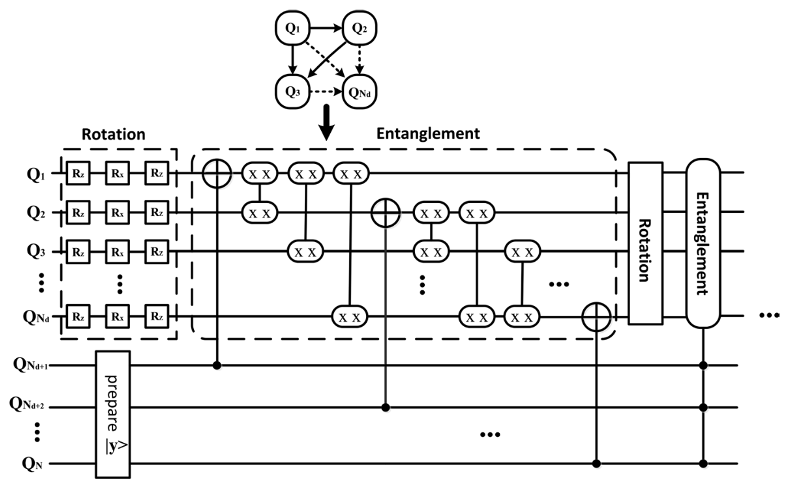}
\caption{QGAN Quantum Circuit Generator \cite{Liu2021}}
\label{fig:17_from_Liu2021}
\end{figure}

The gates $R_{x}$ and $R_{z}$ being defined as:

\begin{equation}
R_{x}(\theta) =    
    \begin{bmatrix}
        cos(\theta/2) & -i sin(\theta/2) \\
        -i sin(\theta/2) & cos(\theta/2)  
    \end{bmatrix}, 
R_{z}(\theta) =    
    \begin{bmatrix}
        e^{-i\theta/2} & 0 \\
        0 & e^{i\theta/2)} 
    \end{bmatrix}
\end{equation}

The $XX$ in [fig:\ref{fig:17_from_Liu2021}] represents an operation involving two qubits, where one is the control qubit, and the other is the target qubit.

The $N_{c}$ qubits are only responsible for passing the conditional information to the other $N_{d}$ qubits and continuing to pass the conditional information to the discriminator in post-processing.

\subsubsection{Recurrent quantum neural networks (RQNN)}

\begin{figure}[!htbp]
    \centering
    \subfloat[QNN]{\includegraphics[width=0.3\linewidth]{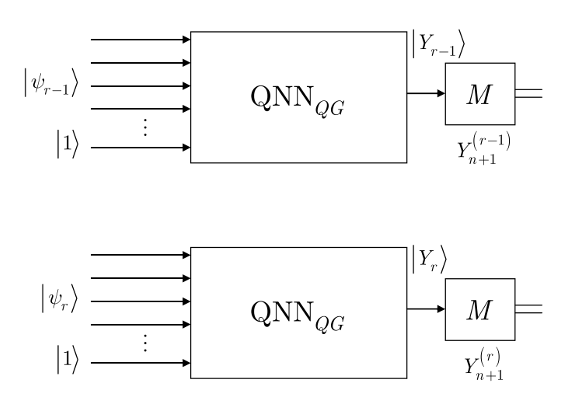}\label{fig:19_from_Gyongyosi2019_a}}
    \subfloat[RQNN]{\includegraphics[width=0.3\linewidth]{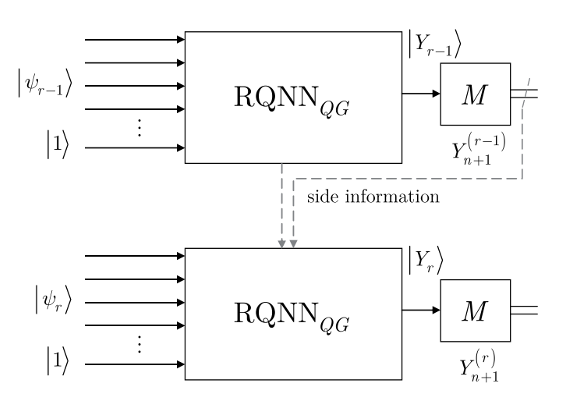}\label{fig:19_from_Gyongyosi2019_b}}
    \hfill
    \caption{QNN y RQNN \cite{Gyongyosi2019}}   
    \label{fig:19_from_Gyongyosi2019}
\end{figure}

A comparison of the models of a $QNN_{QG}$ and $RQNN_{QG}$ for an $(r-1)$-th and $r$-th measurement round [fig:\ref{fig:19_from_Gyongyosi2019}] is performed in \cite{Gyongyosi2019}. The input values are represented by $\ket{\psi_{r-1}}\ket{1}$ and $\ket{\psi_{r}} \ket{1}$, while the output values are denoted by $\ket{Y_{r-1}}$ and $\ket{Y_{r}}$. The output of the measurement operator $M$ in the $(r-1)$-th and $r$-th measurement rounds is denoted by $\ket{Y^{(r-1)}_{n+1}}$ and $\ket{Y^{(r)}_{n+1}}$. 

In [fig:\ref{fig:19_from_Gyongyosi2019_a}], the structure of a $QNN_{QG}$ for an $(r-1)$-th and $r$-th measurement round is depicted, while in [fig:\ref{fig:19_from_Gyongyosi2019_b}], the structure of an $RQNN_{QG}$ is illustrated. 

The advantage of this type of network is that information about gate parameters, and measurement results are transmitted from the $(r-1)$-th measurement round (represented by the dashed grey arrows) to the $r$-th measurement round.

Another implementation of RQNN is presented in \cite{bausch2020recurrent}, where the authors proved that, even if the transformations in quantum circuits -with gates and measurements- will necessarily be linear operations, nonlinear behaviors may occur during these transformations. For example, in 

\begin{equation}
    \textbf{R}(\theta) = \exp
                \begin{pmatrix}
                {i\theta} 
                \begin{pmatrix}
                0 & -i\\
                i &  0 
                \end{pmatrix}
                \end{pmatrix}
            =
                \begin{pmatrix}
                \cos\theta &  \sin\theta\\
                -\sin\theta &  \cos\theta 
                \end{pmatrix} 
                ,
\end{equation}

we can see that the rotation transformation is a linear operator; meanwhile, the amplitudes of the state, $\cos{\theta}$ and $\sin{\theta}$ depend non-linearly on the angle $\theta$.

\subsubsection{Quantum recurrent encoding-decoding neural networks (QREDNN)}

In the quantum recurrent encoder-decoder neural network (QREDNNN) \cite{CHEN2020105863}, the attention mechanism is used to reconstruct the encoder and decoder of the QREDNNN simultaneously. With this mechanism, the QREDNNN can suppress the interference of redundant information, obtaining better nonlinear approximation capability. 

On the other hand, the quantum neuron is used to construct a new quantum recurrent unit (QGRU), which is proposed in the same paper. The activation values and weights are represented by quantum rotation matrices. The QGRU has a large number of multiple attractors, so it can replace the traditional recurrent unit of the encoder and decoder and improve the generalization ability and response speed of QREDNN. 

\begin{figure}[!htbp]
\centering
\includegraphics[width=0.4\textwidth]{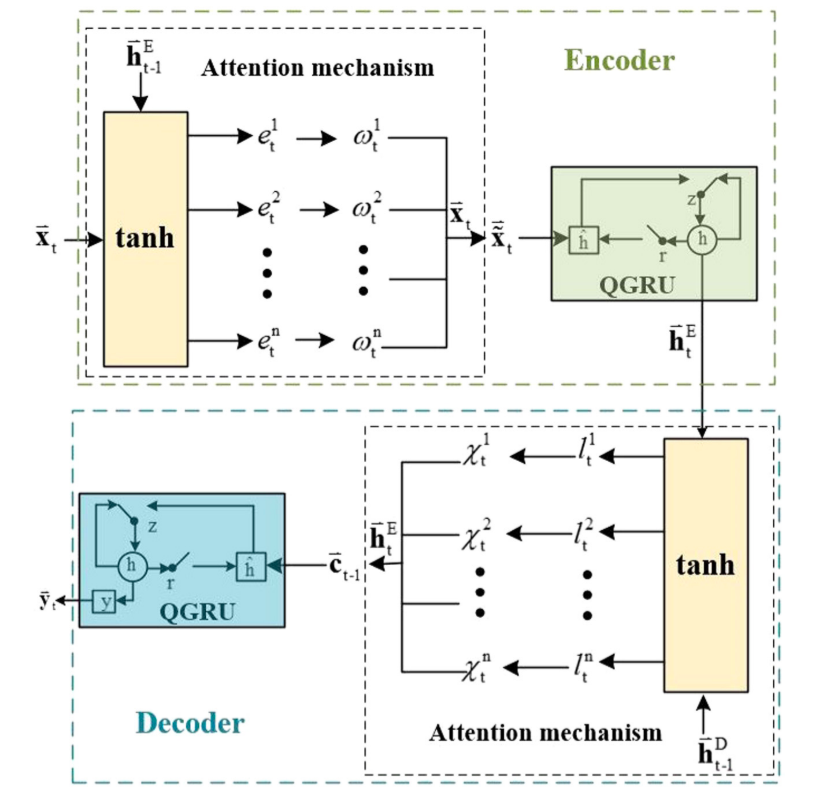}
\caption{Quantum Recurrent Encoding-Decoding Neural Network \cite{CHEN2020105863}}
\label{fig:20_from_CHEN2020105863}
\end{figure}

In addition, the Levenberg-Marquardt (LM) algorithm \cite{More1978} is introduced to improve the update rates of the rotation angles of quantum rotation matrices and the attention parameters of QREDNNN [fig:\ref{fig:20_from_CHEN2020105863}].

\subsubsection{Quantum Vision Transformers}
\begin{figure}[!htbp]
\centering
\includegraphics[width=0.6\textwidth]{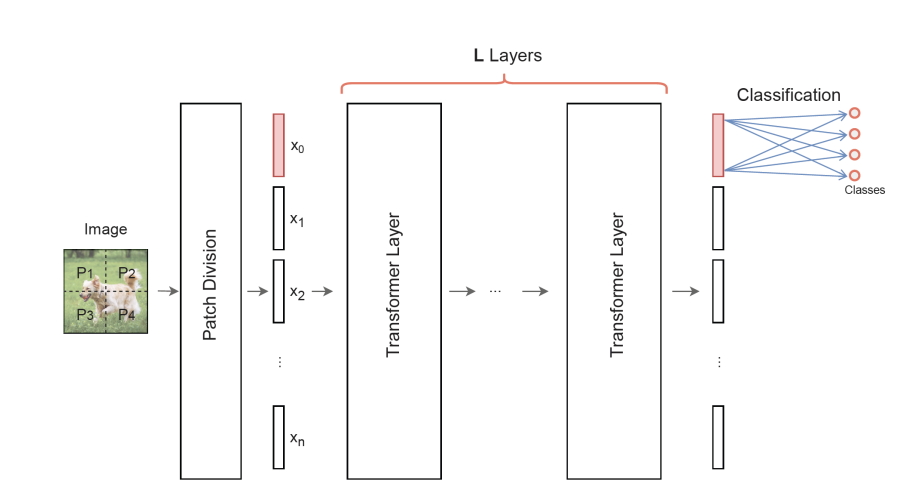}
\caption{Single Transformer Layer \cite{cherrat2022quantum}}
\label{fig:24_from_cherrat2022quantum}
\end{figure}

In \cite{cherrat2022quantum}, the authors designed the architecture for the Quantum Vision Transformers. First, the network decomposes the image into patches and pre-processes the set of patches to map each one into a vector. After that, the feature extraction is performed using a transformer layer, and after that, a normalization layer is applied, and the attention mechanism is. When this is done, another normalization layer is applied, followed by Multi-Layer Perceptron (MLP) [fig:\ref{fig:24_from_cherrat2022quantum}].

To conclude, another important aspect in \cite{cherrat2022quantum} is the usage of the \textit{amplitude encoding} technique, which uses the classical scalar component of the data as amplitudes of a quantum state made of \textit{d} qubits.

\subsubsection{Quantum Monte Carlo}

The authors in \cite{SKAVYSH2023104680} were the first to apply the quantum Monte Carlo algorithm to problems in economics; they began with a Gaussian sampling and, after that, they made stress testing of banks and solved DSGE models with deep learning.

They define two unitary operators to develop a quantum alternative for the Monte Carlo algorithm \cite{monte_carlo}, $A$ for modeling function $p(\cdot)$ and unitary $R$ for modeling function $f(\cdot)$ in the classical equation given by

\begin{equation}
    \mu \coloneqq E[f(x)] = \int_{\mathbb{R}^{d}} p(x)f(x) dx. 
\end{equation}

First, defines $A$ considering a register of $m$ qubits and the initial state $\ket{0}^{\bigotimes m}$

\begin{equation}
    A\ket{0}^{\bigotimes m} =  \sum_{i \in x} \sqrt{p(i}\ket{i} 
\end{equation}

After that, the $R$ operator plus an additional ancilla qubit is defined

\begin{equation}
    R\ket{i}\ket{0} =  \ket{i}\left(\sqrt{1 -f(i)}\ket{0} + \sqrt{f(i)}\ket{1}\right)
\end{equation}

And finally, the output state as 

\begin{equation}
  \ket{\chi}  \coloneqq F\ket{0}^{\bigotimes m+1}
  = \sum_{i \in X} \sqrt{p(i)} \ket{i} \bigotimes \left( \sqrt{1 - f(i)}\ket{0} + \sqrt{f(i)}\ket{1} \right)
\end{equation}

\subsection{Application Domains and Metrics}
Regarding RQ3, the different applications of the models explained in the previous section are presented, distinguishing between applications with a classical purpose, i.e., those that help improve classical processes, and applications with a quantum purpose. In addition, to answer RQ4, we searched and discussed the most relevant metric. Finally, we show a comparison between the selected metric and the accuracy with the different papers and datasets in \hyperref[tab:TablaMNIST]{Table 3}.

\subsubsection{Classic}
Classical purpose applications include classification, in particular image classification, natural language processing (NLP), or applications aimed at improving the process of machine learning itself.

Among the classification applications, we find in \cite{suzuki2021natural} both the resolution of the NARMA task \cite{Appeltant2011}, which is a benchmark test used to evaluate the performance of the model for temporal information processing and the classification of different objects based on the sensor data obtained when grasping them. In addition, in \cite{Abdel-Aty2020}, the Zidan model is presented; this model is used to solve problems based on the degree of entanglement between two auxiliary qubits. Furthermore, this computational model proposes an algorithm that distinguishes the most probable class between two class labels (0 or 1), when an incomplete input is introduced to the proposed model. Regarding sequence memorization and finding structures in a string chain in \cite{bausch2020recurrent} RQNN are used to achieve these tasks. Finally, the model proposed in \cite{Lukac2018} allows classifying and distinguishing in Boolean AND, OR, XOR, and NAND functions.

\begin{figure}[!htbp]
\centering
\includegraphics[width=0.4\textwidth]{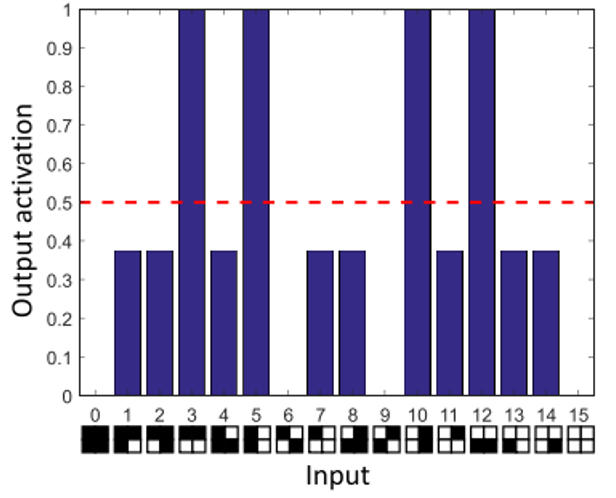}
\caption{Classification of 2x2 pixel images \cite{Tacchino2020}}
\label{fig:21_from_Tacchino2020}
\end{figure}

\begin{figure}[!htbp]
\centering
\includegraphics[width=0.7\textwidth]{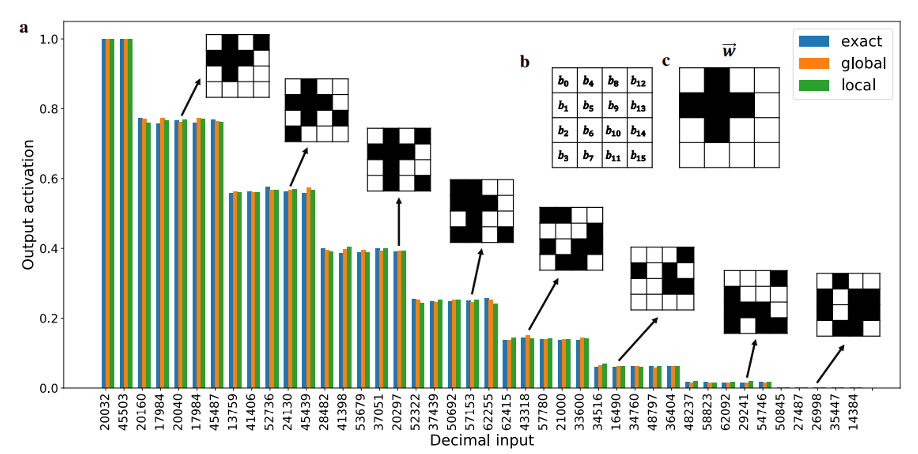}
\caption{Classification of 4x4 pixel images \cite{Tacchino2020}}
\label{fig:22_from_Tacchino_2021}
\end{figure}

Regarding image classification models, we find pattern recognition in images \cite{Tacchino2020}; this model allows classifying if there are straight lines in images of 2x2 pixels, either horizontal or vertical lines [fig:\ref{fig:21_from_Tacchino2020}]. The same author also presents in \cite{Tacchino_2021} a classification model for 4x4 pixel images [fig:\ref{fig:22_from_Tacchino_2021}].

On the other hand, classification models which are trained with well-known datasets such as the MNIST \cite{Lecun1998}, images of handwritten digits from 0 to 9, the IRIS dataset \cite{Fisher1936} that represents by four features three species of the Iris flower, the Cancer Imaging Archive (TCIA) \cite{Clark2013} hosts a large archive of medical images of cancer, the German Traffic Sign Recognition Benchmark (GTSRB) \cite{Stallkamp2012} contains 43 classes of traffic signs, the Banknote Authentication (BNA) \cite{Lohweg2012} includes images taken from forged banknote-like specimens. In addition, the Wireless Indoor Localization dataset (WIL) \cite{Bhatt2017} holds wifi signal strength observed on a smartphone, the Plane Point dataset \cite{Li2020c} contains two types of circles, the Cats vs. Dogs dataset \cite{Parkhi2012} includes different breeds of cats and dogs, CIFAR-10 \cite{krizhevsky2009cifar} contains a collection of images commonly used in machine learning like ten different classes represent airplanes, cars, birds, dogs, horses, ships, etc. Other datasets used in the reviewed papers are MedMNIST, a collection of 12 pre-processed open medical image datasets  \cite{Yang_2021}, and Fashion-MNIST \cite{xiao2017_online} is a group of 28x28 grayscale images, associated with a label from 10 classes.


\begin{sidewaystable}
\begin{longtable}
		{|p{2,5cm}  p{0,4cm}  p{0,4cm}  p{0,4cm}  p{0,4cm}  p{0,4cm}  p{0,4cm}  p{0,4cm}  p{0,4cm}  p{0,4cm}  p{0,4cm} p{0,4cm} p{0,4cm} p{0,4cm} p{0,4cm} p{0,4cm} p{0,4cm} |} \hline
		   \textbf{Dataset} & \cite{HUANG202189} & \cite{Li2020b} & \cite{kerenidis2021classical} & \cite{thumwanit2021trainable} 
		   & \cite{Wall_2021} & \cite{Konar2021} & \cite{Chalumuri2021} & \cite{Li2020c} & \cite{Chen2021} & \cite{tomesh2020coreset} &
            \cite{Du_2022} &
            \cite{bausch2020recurrent} &
            \cite{Huang_2021} &
            \cite{Alam_2022} &
            \cite{Reya_2022} &
            \cite{cherrat2022quantum}       
            \\ \hline 
            \textbf{MNIST}                  & X & X & X & X & X &   &   &   &   &   & X & X & X & X & X &  \\ \hline
            \textbf{TCIA}                   &   & X &   &   &   & X &   &   &   &   &   &   &   &   &  &      \\ \hline
            \textbf{GTSRB}                  &   & X &   &   &   &   &   &   &   &   &   &   &   &   &   &    \\ \hline
            \textbf{IRIS}                   & X &   &   &   &   &   & X & X &   &   &   &   &   & X  &  &    \\ \hline
            \textbf{BNA}                    &   &   &   &   &   &   & X &   &   &   &   &   &   &   &  &     \\ \hline
            \textbf{WIL}                    &   &   &   &   &   &   & X &   &   &   &   &   &   &   &  &     \\ \hline
            \textbf{Plane Point}            &   &   &   &   &   &   &   & X &   &   &   &   &   &   &  &     \\ \hline
            \textbf{Cats vs Dogs}           &   &   &   &   &   &   &   &   & X &   &   &   &   &   &  &     \\ \hline
            \textbf{CIFAR-10}               &   &   &   &   &   &   &   &   & X & X &   &   &   &   &  &    \\ \hline
            \textbf{MedMNIST}               &   &   &   &   &   &   &   &   & X & X &   &   &   &   &   & X  \\ \hline
            \textbf{Fashion-MNIST}          &   &   &   &   &   &   &   &   & X & X &   &   &   &  X &   &  \\ \hline            
    \caption{Common Datasets}
	\label{tab:Tabla1Columnas}
\end{longtable}
\end{sidewaystable}

In order to measure the success rate of the algorithms, the accuracy metric is used in the majority of the papers; this metric is useful because it is possible to compare it with the results obtained using classical algorithms in other articles, given the open access to the different datasets. In the following tables, we show the results of the accuracy metrics obtained for these datasets.

\begin{sidewaystable}
\begin{longtable}
		{|p{2,5cm} p{2,5cm} p{4,5cm}  p{3,5cm}|} \hline
		   Article & Dataset & Training Precision (\%) & Testing Precision (\%)\\ \hline 
            \cite{HUANG202189}                  & MNIST & 83,36 & \\ \hline
            \cite{Li2020b}                  & MNIST & 99,16 & 98,97\\ \hline
            \cite{kerenidis2021classical}   & MNIST & 98,40 & \\ \hline
            \cite{thumwanit2021trainable}   & MNIST & 91,10 & 91,70\\ \hline
            \cite{Du_2022}   & MNIST &  & 98,20\\ \hline
            \cite{Alam_2022}   & MNIST & 93,60 & 87,50\\ \hline
            \cite{Reya_2022}   & MNIST & 99,69 & \\ \hline    
            \cite{Alam_2022}   & Fashion-MNIST & 97,66 & 93,16\\ \hline            
            \cite{Li2020b}                  & GTSRB & 92,56 & 91,40\\ \hline
            \cite{Konar2021}                  & TCIA & 99,60 & \\ \hline
            \cite{Chalumuri2021}            & IRIS & 87,50 & 92,10\\ \hline
            \cite{Li2020c}                  & IRIS & 93,33 & 96,66\\ \hline 
            \cite{Chalumuri2021}            & BNA & 88,53 & 89,50\\ \hline
            \cite{Chalumuri2021}            & WIL & 89,60 & 91,73\\ \hline
            \cite{Li2020c}                  & Plane Point & 97,07 & 96,82\\ \hline
            \cite{Chen2021}             & Cats vs Dogs & 98,70 & \\ \hline
            \cite{Chen2021}             & CIFAR-10 & 94,05 & \\ \hline
            \cite{cherrat2022quantum}   & PathMNIST &  & 75,50\\ \hline
            \cite{cherrat2022quantum}   & ChestMNIST &  & 94,80\\ \hline
            \cite{cherrat2022quantum}   & DermaMNIST &  & 72,70\\ \hline
            \cite{cherrat2022quantum}   & OCTMNIST &  & 60,80\\ \hline
            \cite{cherrat2022quantum}   & PneumoniaMNIST &  & 90,20\\ \hline
            \cite{cherrat2022quantum}   & RetinaMNIST &  & 54,80\\ \hline
            \cite{cherrat2022quantum}   & BreastMNIST &  & 83,30\\ \hline
            \cite{cherrat2022quantum}   & BloodMNIST &  & 88,80\\ \hline
            \cite{cherrat2022quantum}   & TissueMNIST &  & 59,60\\ \hline
            \cite{cherrat2022quantum}   & OrganAMNIST &  & 77,00\\ \hline
            \cite{cherrat2022quantum}   & OrganCMNIST &  & 78,70\\ \hline
            \cite{cherrat2022quantum}   & OrganSMNIST &  & 62,00\\ \hline
            
    \caption{Datasets Metrics}
	\label{tab:TablaMNIST}
\end{longtable}
\end{sidewaystable}

In addition, other classical applications are the implementation of a generic mathematical model that takes advantage of quantum parallelism to speed up machine learning algorithms \cite{Dasari2020}; this model is applied in high dimensional vector spaces or the implementation of an architecture to simulate the calculation of an arbitrary non-linear function \cite{DePaulaNeto2020}. Furthermore, an example of training using OpenAI Gym, a toolkit for developing and comparing machine learning algorithms, is to solve the frozen lake game, where the player moves across a board with tiles, and the model must learn not to fall into the tiles with water on them \cite{Chen2020}. Also, we can find a model able to forecast 24-hour-ahead wind speeds in \cite{Hong2023c} and articles about high entropy alloys \cite{Brown2023a}. Among the applications related to the physics of the Large Hadron Collider (LHC), we find a study that applies a quantum support vector machine (QSVM-Kernel) to try to probe the coupling of the Higgs boson to the top quark, the heaviest known fermion. \cite{wu2021application}. 

To conclude, we elaborate a classification of real-world problems and papers, which the structure presented in \cite{JADHAV20232612}.

\begin{sidewaystable}
\begin{longtable}
		{|p{4,5cm}  p{0,3cm}  p{0,3cm}  p{0,3cm}  p{0,3cm}  p{0,3cm}  p{0,3cm}  p{0,3cm}  p{0,3cm} p{0,3cm}  p{0,3cm}  p{0,3cm} p{0,3cm}  p{0,3cm}  p{0,3cm} p{0,3cm}  p{0,3cm}  p{0,3cm} p{0,3cm}  p{0,3cm} |} \hline
		   \textbf{Algorithm} & 
           \cite{Tacchino2020} & \cite{Tacchino_2021} & \cite{HUANG202189} &\cite{Li2020b} & \cite{kerenidis2021classical} & \cite{thumwanit2021trainable} & \cite{Chalumuri2021} & \cite{Alam_2022} &\cite{Yan2020} & \cite{tomesh2020coreset} & \cite{HUANG202189} & \cite{Li2020c} & 
              \cite{9627527} & \cite{coecke2020foundations} & \cite{9652465} & 
		   \cite{Shenoy2020} & 
		   \cite{Bhatt2017} & \cite{ASHWIN2023108565} & 
                \cite{wu2021application} \\ \hline

            \textbf{Image Classification}                  & X  &  X & X  & X  & X  &  X &  X & X  &  X & X  & X  &  X &   &   &   &   &   &   &     \\ \hline
            \textbf{Natural language Processing}                    &   &  &   &   &   &   &   &   &   &   &   &   &   &   &   &   &   &   &   \\ \hline
            \textbf{Reinforcement Learning}                    &   &   &   &   &   &   &   &   &   &   &   &   &   & X  &  X &  X &   &   &   \\ \hline
            \textbf{Wireless Communication}                    &   &   &   &   &   &   &   &   &   &   &   &   &   &   &   &   & X  &  X &   \\ \hline
            \textbf{Collider phenomenology}                    &   &   &   &   &   &   &   &   &   &   &   &   &   &   &   &   &   &   &  X  \\ \hline            
    \caption{Relationship Between Paper and Solved Problem}
	\label{tab:Tabla3Columnas}
\end{longtable}
\end{sidewaystable}

\subsubsection{Quantum}

The quantum applications include areas related to quantum circuits like data encoding, compilers or the detection of entanglement, identification of faulty gates, and quantum machine learning. The last one has subareas such as feature reduction or quantum natural language processing.

In \cite{Larose2020}, the data encodings for binary quantum classification and their properties in measurements obtained with and without noise are studied. In this data preparation stage, the authors in \cite{Selig2023} proposed an NN called DeepQPrep, that is able to generate quantum circuits for state preparation. Another quantum circuit application is a compiler that can replace the Qiskit compiler, which uses low-depth, IBM Quantum-compatible quantum circuits that generate Greenberger-Horne-Zeilinger (GHZ) interleaved states \cite{Ferrari2018}. Regarding quantum resource reduction, a sorting method designed for quantum processing units that allow working with a smaller number of quantum resources but without any decrease in efficiency is presented in \cite{Sadowski2018}. This is achieved through a classification protocol with the information distributed among several agents. In addition, in \cite{10258300}, try to construct streamlined circuits using minimal qubits and entanglement gates to evade barren plateau phenomena and reduce computational times. Finally, in \cite{Qiu2019}, discrete-variable and continuous-variable quantum neural networks are used to detect entanglement in different circuits. Regarding the error correction problem, the authors in \cite{Convy_2022} propose an ML algorithm using a recurrent neural network to identify bit-flip errors from continuous noisy syndrome measurements.
Moreover, we find techniques that attempt to clone quantum states. Since the no-cloning theorem states that creating an identical copy of an arbitrary unknown quantum state is impossible, techniques are being developed to clone unknown states with high fidelities rather than exact copies. For example, in \cite{Shenoy2020}, the authors attempted to clone an unknown state in IBM's QASM simulator using a quantum reinforcement learning protocol instead of the usual cloning method, quantum tomography. Finally, the reconstruction of the distribution of measurement results in a quantum circuit is presented in \cite{Situ2020}; the authors used a variational autoencoder (VAE). Also, they determined that states with higher entanglement are more challenging to learn, and due to current implementations of quantum logic gates can be very faulty and introduce errors, a model is presented that identifies such faulty gates \cite{LaBorde2020}. Concerning the task of finding an interpretation or explanation for quantum circuits, the authors in \cite{heese2023explaining} assign a Shapley Value \cite{shapley1952, aumann1974} to every gate of the circuit and consider that a value function that is determined by the measurement results. With this method, they are able to evaluate the quality of different ansatz.

On the other hand, we have quantum machine learning applications, like the gradient calculation in neural networks in circuits with noise \cite{Patterson2021}, whose authors find that for the computation of the required gradient in a noisy device, a quantum circuit with a large number of parameters is disadvantageous. This limitation is overcome by introducing a smaller circuit, and the QNN performs well at current quantum hardware noise levels. Also, it is shown that a model with fewer parameters present in the output classifiers of quantum generative adversarial networks (GANs) facilitates discrimination. As well as that, noise in quantum devices increases the overlap between states as the gates of the circuit are applied. In addition, in \cite{Mari_2020}, the authors proposed a transfer learning method between classical-quantum models, quantum-classical models, and quantum-quantum models.
Moreover, in \cite{Chakraborty2020}, a new quantum feature subset selection technique is presented that helps to reduce the dimension of a dataset; this is achieved using quantum parallel amplitude estimation and quantum amplitude amplification techniques. Finally, we find the conceptual and mathematical foundations for quantum natural language processing (QNLP) are provided in \cite{coecke2020foundations} and a benchmark to evaluate the performance of matrix product state (MPS) in quantum assisted machine learning (QAML) models \cite{Wall_2021}.

Another important topic is Quantum Error Correction, in \cite{9502081, Qin_2022} compares different quantum error correction techniques in Quantum state tomography (QST), for example, Measurement error mitigation (MEM) [27] which can mitigate the errors of a noisy measurement device, Zero-noise extrapolation (ZNE) which extrapolates the noise-free expectation value of an observable M[20], Least square error mitigation (LSEM) [29] that estimates the expectation of M at multiple random noise levels and the application of Neural network models for reduce state preparation and measurements (SPAM) errors with a neural network (NN)[15]. Also, approach based on Coherent Pauli Checks (CPCs) \cite{berg2022singleshot} is able detect errors in a Clifford circuit by verifying commutation rules between random Pauli-type check operators and the considered circuit and in \cite{ravi2021vaqem} the authors proposed Variational Approach to Quantum Error Mitigation (VAQEM), which dynamically tailors existing error mitigation techniques to the VQAs by tuning specific features of these mitigation techniques. In \cite{jose2022error}, authors try to mitigate the impact of the quantum gate noise on the trainability of parameterized quantum circuits (PQCs). The authors analyze different impacts like shot noise or gate noise and introduce a quasi-probabilistic error mitigation (QEM) for evaluating the stochastic gradient estimator on the noisy PQC. To conclude, there exist metrics for the measurement of error mitigation; in \cite{saki2023hypothesis}, they combine some mitigation error methods such as measurement error mitigation (MEM), randomized compiling (RC), zero noise extrapolation (ZNE), and dynamical decoupling (DD) for the metric creation.

\subsection{Devices}

In the case of RQ5, we introduce the different hardware environments used in the papers and their particularities. Implementing quantum devices is necessary because the same quantum circuit with different implementations can produce different results. For example, depending on the implementation, the relaxation time (T1), the time it takes for the qubit to decay from state $\ket{1}$ to $\ket{0}$, or the decoherence time (T2), the time at which the qubit loses its coherence, are different. Other factors that can be affected are the state preparation, the efficiency in applying gates, and the readout results.

One of the first companies to develop quantum devices was IBM. Since 2016, it has offered open access to its experimental quantum computing platform, called IBM Quantum. All the devices are based on superconducting qubits that are insensitive to charge noise and have a decoherence time of $\simeq100$ us \cite{Ferrari2018}. IBM Quantum \url{https://quantum-computing.ibm.com/} is an experimental platform for testing quantum algorithms. A visual web tool allows for composing quantum circuits and running simulated or real experiments. Moreover, circuits can be designed using the OpenQASM language \cite{Cross2017}, and experiments can be defined and run using the Qiskit Python library \cite{Qiskit}. Not all physical gates are implemented directly in the devices, but a transpilation process takes place to convert the programmed gates into physical gates.

Rigetti systems are powered by superconducting qubit-based quantum processors, in which qubits are coupled to a linear superconducting resonator for readout. Their principal advantages are fast gate times and fast program execution times, making them ideally suited to NISQ-era applications and the requirements for quantum error correction and fault-tolerant quantum computing. Among the tools offered by Rigetti are PyQuil \cite{smith2016practical} as a Python library, which allows both programming and compiling quantum circuits. To compile these circuits, the Quilc \url{https://github.com/quil-lang/quilc} compiler is used. This compiler accepts not only PyQuil as input but also Qiskit, Cirq \cite{cirq_developers_2021} or OpenQASM language and also allows execution on non-Rigetti quantum processors. For the simulation of the execution of these circuits, Rigetti provides access to the Quantum Virtual Machine QVM \url{https://pyquil-docs.rigetti.com/en/1.9/qvm.html}, the open-source implementation of a quantum computing simulator.

TensorFlow Quantum (TFQ) \cite{broughton2021tensorflow} is a quantum machine learning library used for prototyping classical-quantum hybrid models. It uses the Cirq language, and these models are compatible with existing TensorFlow APIs. In addition, high-performance quantum circuit simulators are provided to run them. Amazon Braket \url{https://aws.amazon.com/es/braket/} provides access to quantum processing unit (QPU) devices from different vendors such as D-Wave \url{https://www.dwavesys.com/}, IonQ \url{https://ionq.com/} and Rigetti \url{https://www.rigetti.com/}, as well as three proprietary simulators (SV1 DM1 and TN1). Matlab \url{https://matlab.mathworks.com/} provides a library that implements the simulation of a universal quantum computer. It allows the user to simulate quantum algorithms before implementing them on real quantum computers.

In \cite{OQuinn2020} a review of different platforms that offer quantum simulation services, such as the IBM Quantum's simulator, called \textit{ibmq\_qasm\_simulator} \cite{wu2021application} \cite{Yan2020} \cite{Li2020b} in various articles, as well as the use of different physical devices such as \textit{ibmqx2} \cite{Kathuria2020} \cite{Dash2020} \cite{Gokhale2020}, \textit{ibmqx4} \cite{Gokhale2020}, \textit{ibmq 16 melbourne} \cite{Kathuria2020} \cite{Chalumuri2021} \cite{suzuki2021natural}, \textit{ibmq\_rome} \cite{Yan2020}, \textit{ibmq\_santiago} \cite{Yan2020}, \textit{ibmq\_paris} \cite{wu2021application} and \textit{ibmq\_toronto} \cite{suzuki2021natural} or Rigetti devices \cite{Shenoy2020} \cite{Romero2021} \cite{Anand2021}, Google Tensorflow Quantum Simulator \cite{wu2021application}, Amazon Braket Local Simulator \cite{wu2021application} or the Matlab simulator \cite{Lee2016} \cite{Lukac2018}.

\section{Discussion}
Among the algorithms used, most of them used the qubit as a minimum unit to use them in the circuit definition and implementation. However, an alternative is shown in \cite{Konar2021} where the qutrit is presented as a helpful alternative in specific tasks. Moreover, we observe many techniques derived from neural networks, such as convolutional, orthogonal, feedback, deep, and autoencoders. The use of quantum neurons and the quantum circuit implemented -designed using rotational gates and auxiliary qubits to perform operations- are the standard features in the different varieties of QNNs. This development of quantum neural networks stems from the fact that they allow their correct execution and can provide a quantum advantage in noisy quantum computers. However, other algorithms like the QSVM give rise to a $2^{N}$ dimensional feature space (N is the number of qubits), which potential will be achieved after the NISQ era.
In contrast to the neural networks, we detect few articles in other disciplines of machine learning, like natural language processing or linear regression. The articles found in these two areas establish the basis to define and implement these new emerging fields. Also, most of these machine learning models are hybrid implementations, i.e., the models consist of both classical and quantum layers. In summary, hybrid implementations of classical machine learning models appear, such as the k-nearest neighbor model or support vector machines. The quantum algorithm is applied in a specific task of the process, i.e., calculating the distances between the vectors and finding the separation of the hyperplane, respectively. We find many hybrid implementations where the quantum algorithms are applied to improve a specific step of the classical process. Moreover, a review of the Variational Quantum Eigensolver (VQE) is done in \cite{Tilly_2022}. This algorithm is used to model complex wavefunctions in polynomial time and present some degree of resilience to the noise in the quantum hardware. However, a large number of measurements are required to perform the optimization process.

In classical applications, one of the main lines of research is focused on classification. Although some papers focus on label classification, most of them focus on image classification. Image classification algorithms use known datasets such as MNIST or IRIS, or 2x2 or 4x4 pixel images, consisting of a reduced number of categories to differentiate (due to the low number of qubits currently available in quantum devices). Even if these problems are already solved in classical machine learning, they achieve comparable results in terms of accuracy. Also, \cite{Chakraborty2020} shows a comparison in terms of CPU computational time, and the hybrid approach can complete the training process in a shorter time than other classical approaches. However, some articles already try to take advantage of quantum advantages to achieve results that are not achievable with classical algorithms, such as the use of quantum parallelism to improve the speed of machine learning algorithms \cite{Dasari2020}. The latter articles are necessary to demonstrate the quantum advantage over classical computing. On the other hand, there are also newly discovered fields like QNLP, such as \cite{coecke2020foundations}, where the authors establish a possible basis for further development and implementation.

In quantum applications, we notice a large number of articles that aim to improve the process of creating quantum circuits. These articles cover all the processes of creation of the ansatz itself; we highlight \cite{Sadowski2018} an article that can allow NISQ devices, i.e., devices with current technology, to achieve outstanding results by being able to use a small number of quantum resources. Another relevant article is the one that challenges the no-cloning rule in \cite{Shenoy2020} without using quantum tomography. Regarding the specific applications of quantum machine learning, the aim is to improve the process of creating neural networks. The creation of models with quantum applications is driven by the need to improve current noisy quantum computers and their physical limitations. In addition, the authors in \cite{Nguyen_2022} created a platform named Qsun, where a wavefunction represents a quantum register, quantum gates are manipulated directly by updating the amplitude of the wavefunction, and the measurement results rely on probabilities of the wavefunction. Also, this platform has QML algorithms implemented, like quantum linear regression or quantum neural networks.

In terms of devices, we find a wide variety of quantum simulators on different platforms, which allow testing as if they were real quantum devices and even allow the noise of these devices to be simulated. So, the current hardware capacity does not hold back the potential for algorithm and software development. We also find that most platforms offer support or allow translation for the OpenQASM language, which is universal, allowing research and implementation without restrictions on specific private technologies. Quantum devices today do not present a similarity in their hardware; we can observe both devices made with superconductors (IBM or Rigetti) and devices with trapped ions (IonQ). In addition, devices are made with topological cubits, photons, NV center (diamond), neutral atoms, or quantum dots. Each company opts for one or more technologies when it comes to physically implementing the devices. Each one has its advantages and disadvantages, there is no consensus or predominance of one in the literature. Today, access to quantum devices for regular users is provided by large companies through cloud computing with limitations. However, everyday access to them is not yet possible due to the size of the devices, the complexity of implementation, and constant improvement.

To conclude, we want to note some brief comments about the potential threats to validity for this study. To mitigate the bias in certain aspects in the paper, we applied some techniques such as the quality assurance checklist \cite{kitchenham2007guidelines}. Also, to reduce the bias triggered by the number of researchers involved in selecting papers, the first author was the primary reviewer. The other two authors validated the steps of the process. In addition, the public Git repository that contained the results from databases and other resources used in this research can be consulted to reproduce the process.
Regarding the databases used, all the central relevant digital libraries and databases in the field of computer science were included. However, we introduce IBM Quantum Network -a not indexed ensured contrasted quality content- because it contains essential articles not included in the other databases. To ensure the validity of the results and reduce the trustiness of the discussion and conclusions achieved, we have been as rigorous as possible with the process and methodology. For this reason, we have used general terms to perform the query, well-known inclusion and exclusion criteria, and we make available the traceability of the whole process. Furthermore, all the authors have reviewed all the materials and are publicly available. So, we believe that our results deserve enough confidence due to the rigorous methodology process.

\section{Conclusions}
This paper presents a systematic review to identify, classify, and analyze quantum machine learning algorithms and their applications. For this, all the publications related to this knowledge area were examined unbiasedly. During the review process of the papers published between 2017 and 2022, we initially found 5497 papers from 5 different databases. Then, we reduced this amount by removing duplicates and using different criteria and quality assessments. Finally, we got the 94 papers deemed the most relevant for this research. 

Considering the content of the papers, we detected different quantum machine learning designs and implementations. The main trends were related to the neural networks, and we find various quantum networks like the orthogonal, convolutional, feed-forward, or self-supervised. Also noteworthy is the use of linear regressions, quantum amplitude estimation (QAE), variational depth quantum circuit (vVQC), quantum Boltzmann machines (QBM). Regarding classical applications, a large number of papers focus on image classification with well-known datasets such as MNIST. On the other hand, in quantum applications, many of them focus on improving the process of creating, simulating, and executing ansatzs.

During our literature review, we detected the existence of standardization in the encoding of the data and the initialization of the states, but many different implementations and proposals in the creation of oracles and quantum circuits in general. Nevertheless, classical applications try to emulate the problems already solved by classical machine learning, obtaining, in the best case, the same result as the existing classical models.

An improvement in the quantum hardware will support the improvement of the results of the applied techniques. Because many of the applied methods, such as image classification, are limited by the different possible output labels, which are proportional to the number of qubits in the device.





\bibliographystyle{elsarticle-num}
\bibliography{bibliography}







\end{document}